\documentclass[11pt]{article}
\usepackage[margin=1in]{geometry}
\usepackage{fontspec}
\usepackage{amsmath,amssymb,mathrsfs}
\usepackage{longtable,booktabs,array,calc,graphicx,needspace}
\title{Three-body corrections to the Brownian viscosity of hard spheres without hydrodynamic interactions}
\author{Julian Olszewski\textsuperscript{1,‡}, Zijian
Liu\textsuperscript{1,‡} and Jie Xu\textsuperscript{1,*}}
\date{}

\begin{document}
\maketitle

\textsuperscript{1} Department of Mechanical and Industrial Engineering,
University of Illinois Chicago, 842 West Taylor Street, MC 251, Chicago,
IL 60607, USA

\textsuperscript{‡} Julian Olszewski and Zijian Liu contributed equally
to this work.

\textsuperscript{*} Corresponding author: jiexu@uic.edu

ORCID: Julian Olszewski, 0009-0006-2346-8097; Zijian Liu,
0009-0000-6014-1268; Jie Xu, 0000-0001-5765-7431. Electronic mail:
jolsz2@uic.edu, zliu232@uic.edu, jiexu@uic.edu.

Short title: Three-body viscosity corrections

\section{Abstract}\label{abstract}

For Brownian hard spheres without hydrodynamic interactions, we study the part of the contact stress produced by three-body
correlations: the third-order coefficient of the zero-shear Brownian
viscosity. Kirkwood
superposition gives zero for this coefficient: three nonzero pair
contributions cancel exactly. An elementary radial identity exposes the
cancellation, and evaluating closures with the same contact stress makes their
coefficients directly comparable. Although superposition solves the
bulk equation, it leaves a flux error on each shared contact. This error defines a unique finite-energy weak correction whose positive
diffusive dissipation, suitably normalized, is a weak-energy
coefficient. Explicit symmetric trials yield a variational lower bound and
a residual upper bound, and exact integration with outward rational arithmetic proves that the weak-energy coefficient lies between
1.8266948 and 1.8267157. For the physical viscosity coefficient this is a
conditional interval: the two coefficients coincide when the physical
three-body correction equals the weak correction and the remote
remainder of the pair adjoint identity, a surface term at large pair
separation, vanishes. The interval excludes every closure coefficient
compared here, including the zero of superposition. Hydrodynamic interactions
contribute at the same orders in real suspensions, where rheometry
measures total viscosities, so the coefficient is a quantitative benchmark
for theories and simulations of dilute three-body Brownian stress in
this model.

Keywords: contact correlations; diffusion-controlled transport;
nonequilibrium closures; suspension rheology; variational bounds

\section{1. Introduction}\label{introduction}

Predicting the stress needed to drive a colloidal suspension requires
connecting particle interactions, the structure created by flow, and
macroscopic momentum transport. Shear distorts the particle arrangement,
while Brownian motion relaxes that distortion; these processes produce a
structural contribution to stress. Constitutive theories and
particle-based simulations therefore depend on how accurately they
represent correlated particle motion. Precisely characterized
hard-sphere models provide reference problems for testing this
connection {[}1, 2{]}.

A useful benchmark should reveal which stress contributions a closure
retains. Here a closure approximates an unknown distribution using
lower-order distributions. At finite concentration, several density orders contribute to a
measured or simulated viscosity, which can obscure errors in a
particular closure. The first viscosity term that requires three
particles provides a more discriminating test: it isolates the effect of
three-body correlations on the pair response. Resolving this term tests
whether a predicted cancellation follows from the shared contact
constraints and supplies a quantitative target for theories and
numerical solvers of the three-body problem.

We consider hard spheres without hydrodynamic interactions. Each sphere
keeps the mobility and free diffusivity of an isolated sphere and is
carried by the undisturbed shear flow, so the spheres move through the
suspending fluid independently and interact only through the contact force that
prevents overlap. This model isolates the stress
generated by diffusion and exclusion constraints. It is the reference
system in which nonequilibrium closure theories are formulated and
tested {[}3{]} and in which Brownian dynamics simulations without
hydrodynamic interactions are performed {[}4, 5{]}. The excluded
radius \(b\) is half the center-to-center distance at contact. The model
is also useful when \(b\) exceeds the hydrodynamic radius \(a\). At these
larger separations, hydrodynamic coupling can be weaker {[}1, 2{]}.

For this model, the Brownian viscosity, the part of the shear response
carried by contact forces, has the zero-shear expansion {[}1, 2{]}

\begin{equation}\frac{\eta_B}{\eta_s}=\frac{12}{5}\phi^2+c_3(0)\phi^3+O(\phi^4),
\label{eq:m-viscosity-expansion}\end{equation}

where \(\eta_s\) is the viscosity of the suspending fluid and \(\phi\) is the sphere
volume fraction. Each power of \(\phi\) counts the spheres that must
meet: the \(\phi^2\) term comes from pair encounters and has the known
coefficient \(12/5\), and the coefficient \(c_3(0)\) describes how a
third sphere changes that response. Equation
\eqref{eq:m-viscosity-expansion} is the contact-force (Brownian) part of
the shear viscosity of this model; each sphere's own stresslet adds the
Einstein term \((5/2)\phi\) {[}1{]}, which involves no interaction
between spheres. In the classical viscosity expansion of a real hard-sphere suspension,
hydrodynamic interactions add further stress from \(O(\phi^2)\) onward and change
the Brownian part itself {[}6, 7{]}; Section~5.3 compares the two expansions order by order.

Computing \(c_3(0)\) requires the three-body distribution, the joint
spatial distribution of three spheres. The pair-distribution equation
contains this distribution through the force exerted by the third
sphere. Kirkwood
superposition uses the product of the three pair distributions {[}8{]}.
It is exact at equilibrium in the infinite-dilution hard-sphere limit.
Under shear, however, the three pair responses must also satisfy their
shared contact constraints.

Lionberger and Russel developed a systematic connection between
equilibrium integral-equation closures and the microstructure of sheared
suspensions {[}3, 9{]}. Lionberger's thesis obtained a vanishing
third-order coefficient from the low-density hypernetted-chain closure
and presented it as the exact virial coefficient of this model
{[}10, p.~110, Eq. (4.107){]}; it also identified the cancellation between the
rise of the equilibrium contact value and the fall of the pair
perturbation at contact {[}10, Eqs. (A.61), (A.62){]}. At this order that
closure reduces to superposition {[}9, 11{]}. This cancellation provides
the starting benchmark for our analysis. Other theories retain different
parts of the three-body response {[}12--14{]}. Lionberger and Russel
also noted that the conditions under which the irreducible three-body
response can be neglected remain unknown {[}9, p.~406{]}. Formal
linear-response theory includes the full three-body contribution in a
cluster expansion {[}15{]}, but evaluating that response remains the
central difficulty.

Building on these closure theories, we connect that response to a specific flux error in the equal-sphere, zero-shear model without
hydrodynamic interactions. Superposition solves the interior equation at
first order in shear but leaves an error in the contact flux. A sharp trace inequality bounds the weighted
mean square of the contact values of a distortion by its diffusive
dissipation; it yields
a unique finite-energy weak correction, whose normalized energy defines
a weak-energy coefficient \(\Theta\). An exactly integrated trial gives a lower bound on
\(\Theta\), and a residual estimate using the rotational symmetry of
shear and the three-sphere geometry gives an upper bound from a separate
trial. Exact entries and outward rational arithmetic prove
\(1.8266948\le\Theta\le1.8267157\). The identification
\(c_3(0)=\Theta\) holds under two conditions that Proposition~1 states
as assumptions: the physical three-body correction equals the weak
correction, and the remote remainder of the pair adjoint identity, a
surface term at large pair separation, vanishes. The result is therefore a proved enclosure of
\(\Theta\) and a conditional enclosure of \(c_3(0)\). All the closure coefficients
evaluated here with the contact stress lie outside the interval; under
the two conditions, the irreducible three-body response, which
superposition omits, carries the whole of \(c_3(0)\).

Section 2 defines the model and contact stress and collects the
principal symbols in Table~\ref{tab:symbols}. Section 3 derives the
superposition cancellation. Section 4 gives the energy identity, its
conditions and two-sided bounds evaluated from exact entries. Section 5
separates what is proved from what is conditional, explains the
physical mechanism of the positive correction, compares the coefficient
with the classical viscosity expansions and with experiments, and
describes its use as a benchmark. Section 6 states the conclusions. The
Supplementary Material gives the closure reductions, trace and residual
estimates, and exact specifications of the two trial fields, together
with the reductions used to evaluate their functionals.

\section{2. Model and contact stress}\label{model-and-contact-stress}

\subsection{2.1 Diffusion, geometry and
stress}\label{diffusion-geometry-and-stress}

We consider identical spheres of radius \(a\) and free diffusivity
\(D_0\) in simple shear, \(\mathbf u^\infty=\dot\gamma y\mathbf e_x\).
Length and time units are \(a\) and \(a^2/D_0\), so contact is at
dimensionless separation 2. The Stokes-Einstein relation is
\(D_0=k_BT/(6\pi\eta_s a)\), with \(k_BT\) the thermal energy. The
Péclet number \(\mathrm{Pe}=\dot\gamma a^2/D_0\) compares shear with
diffusion. Small \(\mathrm{Pe}\) gives a weak distortion of the
equilibrium distribution; \(\mathrm{Pe}\) of order one makes the two
processes comparable. The equal-radius calculation below has \(b=a\);
Section 5 gives the rescaling for a distinct excluded radius.
Table~\ref{tab:symbols} collects the principal symbols and the places
where they are defined.

Let \(\mathbf r=\mathbf x_2-\mathbf x_1\) and
\(\mathbf s=\mathbf x_3-\mathbf x_1\). The accessible configurations
form

\begin{equation}\Omega=\{(\mathbf r,\mathbf s):|\mathbf r|>2,\ |\mathbf s|>2,\ |\mathbf s-\mathbf r|>2\}.
\label{eq:m-accessible-domain}\end{equation}

\begin{figure}[!htbp]
\centering
\includegraphics[width=1\linewidth]{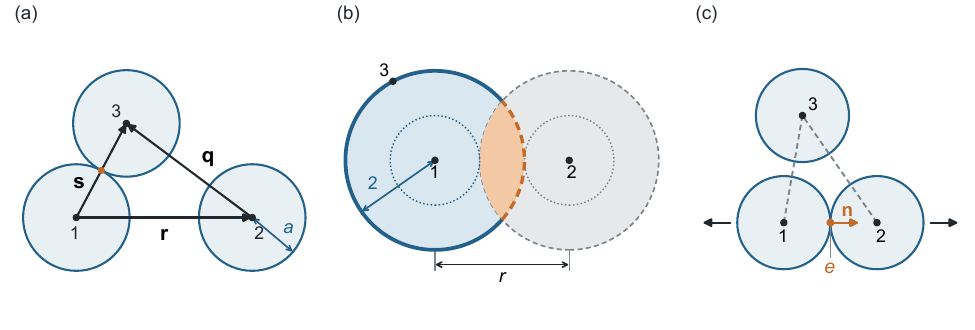}
\caption{Coordinates and contact geometry of three spheres, in units of
the sphere radius \(a\). (a) Relative positions
\(\mathbf r=\mathbf x_2-\mathbf x_1\),
\(\mathbf s=\mathbf x_3-\mathbf x_1\) and
\(\mathbf q=\mathbf s-\mathbf r\). Sphere 3 touches sphere 1 at the
orange point, where \(\mathbf s=2\mathbf n\). (b) Positions available to the
center of sphere 3 for \(2<r<4\), with \(r=|\mathbf r|\), in a plane
through the centers of spheres 1 and 2. The center cannot enter the
discs of radius 2 around sphere 1 (blue) and sphere 2 (gray), which
overlap in the orange lens; the dotted circles are the spheres
themselves. Sphere 3 touches sphere 1 when its center lies on the
boundary of the blue disc: the solid blue arc is accessible and the
dashed orange arc is blocked by sphere 2, which gives the equilibrium
exclusion flux of Section 3. (c) Configuration that generates the
superposition flux error \(e\) of Section 4.1: spheres 1 and 2 touch
along \(\mathbf n\), with sphere 3 nearby. Black arrows show spheres 1
and 2 moving apart; dashed lines mark the pairs 1--3 and
2--3.\label{fig:geometry}}
\end{figure}

Figure~\ref{fig:geometry}(a) shows these relative coordinates,
Fig.~\ref{fig:geometry}(b) the overlap geometry used in Section 3 and
Fig.~\ref{fig:geometry}(c) the contact configuration of Section 4.1.
The free increments of the spheres are independent, while their contact
constraints couple the relative positions. The pair separation has
physical diffusivity \(2D_0\), equal to 2 in the chosen units. Its
infinite-dilution distribution \(g_2\) satisfies
\(2\nabla^2g_2-\mathrm{Pe}\,r_y\partial_xg_2=0\) outside contact, with
zero probability flux into overlap.

\begin{table}[!htbp]
\caption{Principal symbols. Lengths are in units of the sphere radius \(a\) and times in units of \(a^2/D_0\).\label{tab:symbols}}
\centering\small
\renewcommand{\arraystretch}{1.12}
\begin{tabular}{@{}>{\raggedright\arraybackslash}p{0.17\linewidth}>{\raggedright\arraybackslash}p{0.57\linewidth}>{\raggedright\arraybackslash}p{0.2\linewidth}@{}}
\toprule
Symbol & Meaning & Defined in\\
\midrule
\(a,\ b\) & hydrodynamic radius; excluded radius & Secs.~2.1 and 1\\
\(D_0,\ \mathrm{Pe}\) & free-sphere diffusivity; Péclet number \(\dot\gamma a^2/D_0\) & Sec.~2.1\\
\(\dot\gamma,\ k_BT\) & shear rate; thermal energy & Sec.~2.1\\
\(\phi,\ \rho\) & volume fraction; number density \(3\phi/(4\pi)\) & Eq.~\eqref{eq:m-viscosity-expansion}, Sec.~2.1\\
\(\eta_s,\ \eta_B\) & suspending fluid viscosity; Brownian viscosity & Eq.~\eqref{eq:m-viscosity-expansion}\\
\(c_3(0)\) & third-order zero-shear coefficient of \(\eta_B/\eta_s\) & Eqs.~\eqref{eq:m-viscosity-expansion}, \eqref{eq:m-density-contact-coefficient}\\
\(\mathbf r,\ \mathbf s,\ \mathbf q\) & relative positions \(\mathbf x_2-\mathbf x_1\), \(\mathbf x_3-\mathbf x_1\), \(\mathbf s-\mathbf r\) & Fig.~\ref{fig:geometry}(a)\\
\(\Omega,\ \mathbf n\) & accessible three-sphere configurations; unit line of centers at contact & Eqs.~\eqref{eq:m-accessible-domain}, \eqref{eq:m-contact-stress}\\
\(g_2,\ g_3,\ g^{(1)}\) & pair and three-body distributions at infinite dilution; first density correction to the pair distribution & Secs.~2.1, 2.2\\
\(\psi_2,\ H\) & first-order shear response of \(g_2\); its shape \(v_xv_y/|\mathbf v|^5\) & Eq.~\eqref{eq:m-pair-response}\\
\(\psi_3,\ \psi_3^K\) & first-order shear response of \(g_3\); its superposition form & Eqs.~\eqref{eq:m-three-body-adjoint-integral}, \eqref{eq:m-superposition-response}\\
\(w,\ \boldsymbol Q\) & pair adjoint field; third-sphere contact term (\(-2\rho\boldsymbol Q\) is the contact-force part of the pair probability current) & Eq.~\eqref{eq:m-adjoint-flux-identity}\\
\(J_{xy},\ J_1,\ w_0\) & contact projection of \(g^{(1)}\); its first-order coefficient in \(\mathrm{Pe}\); zero-shear adjoint \(\psi_2/8\) & Eqs.~\eqref{eq:m-adjoint-flux-identity}, \eqref{eq:m-three-body-adjoint-integral}\\
\(T_1,\ T_2,\ T_3\) & pair contributions to the superposition coefficient & Eq.~\eqref{eq:m-pair-terms}\\
\(e\) & superposition flux error & Eq.~\eqref{eq:m-contact-flux-error}\\
\(\chi,\ \chi_w\) & physical three-body correction; weak correction & Sec.~4.1, Eq.~\eqref{eq:m-weak-correction}\\
\(\mathcal D,\ \mathcal E\) & Dirichlet form (diffusive dissipation); energy \(\mathcal D/6\) & Eq.~\eqref{eq:m-dirichlet-form}\\
\(\langle f,e\rangle,\ \mathcal H\) & contact pairing; energy space & Eq.~\eqref{eq:m-contact-pairing}, Sec.~4.2\\
\(\mathcal R\) & root-sum-square of the three pair distances & Sec.~4.2\\
\(\Theta,\ K\) & weak-energy coefficient; \(K=81/(32\pi^2)\) & Eq.~\eqref{eq:m-weak-coefficient}, Sec.~4.4\\
\(F(f),\ \eta(v)\) & lower functional of a trial \(f\); squared residual norm of a trial \(v\) & Eqs.~\eqref{eq:m-trial-functional}, \eqref{eq:m-residual-upper-bound}\\
\(\mathcal Lv,\ b_v;\ A_v,\ B_v\) & bulk and contact residuals of a trial \(v\); their weighted norms & Sec.~4.4, Eq.~\eqref{eq:m-upper-residual-norms}\\
\bottomrule
\end{tabular}
\end{table}

The Brownian shear stress follows from the pair distribution at contact
{[}2, 16, 17{]}:

\begin{equation}\frac{\sigma^B_{xy}}{\eta_s\dot\gamma}
=-\frac{27\phi^2}{2\pi\mathrm{Pe}}\oint n_xn_y\,g(2\mathbf n;\phi)\,d\Omega_n.
\label{eq:m-contact-stress}\end{equation}

Here \(\mathbf n\) is the unit line of centers and \(d\Omega_n\) is
solid angle. Supplementary Section S1.1 derives the dimensional stress
and its normalization.
The integral counts contacts weighted by \(n_xn_y\). Shear brings
spheres together along the compressional axis, where \(n_xn_y<0\), and
separates them along the extensional axis, so a sheared suspension has
an excess of compressional contacts, and Eq.~\eqref{eq:m-contact-stress}
turns that excess into a positive viscosity.
Write \(g=g_2+\rho g^{(1)}+O(\rho^2)\), where
\(\rho=3\phi/(4\pi)\) is the dimensionless number density. The
coefficient \(g^{(1)}\) is the first density correction to the pair
distribution, and

\begin{equation}c_3(\mathrm{Pe})=-\frac{81}{8\pi^2\mathrm{Pe}}
\oint n_xn_y\,g^{(1)}(2\mathbf n)\,d\Omega_n.
\label{eq:m-density-contact-coefficient}\end{equation}

Thus \(c_3\) measures how a third sphere changes the shear-weighted
contact probability of a pair.
We use \eqref{eq:m-contact-stress} for every closure comparison.
Supplementary Section S1 derives the changes produced by the
potential-of-mean-force stress expressions used in some earlier
theories.

\subsection{2.2 Pair hierarchy and adjoint
representation}\label{pair-hierarchy-and-adjoint-representation}

The first density correction to the pair distribution is driven by
contact with a third sphere. A pair adjoint identity gives its shear projection without
an explicit solution for the entire density correction. Let \(w\) solve \(2\nabla^2w+\mathrm{Pe}\,r_y\partial_xw=0\)
outside pair contact, with \(2\partial_rw=n_xn_y/4\) at \(r=2\) and
\(w\to0\) at infinity. If the remote remainder, a surface term at large
separation, vanishes, the identity is

\begin{equation}J_{xy}=\oint n_xn_y\,g^{(1)}(2\mathbf n)\,d\Omega_n
=2\int_{|\mathbf r|>2}\boldsymbol Q\cdot\nabla w\,d^3r,
\quad
\boldsymbol Q=-4\oint_{|2\mathbf n-\mathbf r|>2}g_3(\mathbf r,2\mathbf n)\,\mathbf n\,d\Omega_n.
\label{eq:m-adjoint-flux-identity}\end{equation}

Here \(g_3\) is the infinite-dilution three-body distribution and the
factor 4 converts solid angle to contact area. Supplementary Section S1
derives the pair equation, the adjoint normalization and the remote
remainder.
In the pair equation at first order in density, \(-2\boldsymbol Q\) is
the part of the probability current of the pair separation produced by
contact forces from a third sphere, integrated over the third sphere's
accessible contact positions. The adjoint field \(w\) measures how
strongly a disturbance at separation \(\mathbf r\) changes the
shear-weighted contact density of the pair. Equation
\eqref{eq:m-adjoint-flux-identity} therefore expresses the contact projection \(J_{xy}\) as a volume integral of this contact-force current, weighted
by its influence on the stress.

At first order in shear, \(g_2=1+\mathrm{Pe}\,\psi_2+\cdots\), where

\begin{equation}\psi_2(\mathbf v)=-\frac{16}{3}H(\mathbf v),\qquad
H(\mathbf v)=\frac{v_xv_y}{|\mathbf v|^5}.
\label{eq:m-pair-response}\end{equation}

Its contact value \(\psi_2(2\mathbf n)=-2n_xn_y/3\) is the excess of
compressional contacts described after Eq.~\eqref{eq:m-contact-stress};
inserting it in Eq.~\eqref{eq:m-contact-stress} recovers the pair coefficient \(12/5\). The zero-shear adjoint is
\(w_0=\psi_2/8\). Writing \(g_3=1+\mathrm{Pe}\,\psi_3+\cdots\) gives

\begin{equation}c_3(0)=-\frac{81}{8\pi^2}J_1,\qquad
J_1=-8\int_{|\mathbf r|>2}d^3r\oint_{\mathrm{acc}}
\psi_3(\mathbf r,2\mathbf n)\,\mathbf n\cdot\nabla w_0(\mathbf r)\,d\Omega_n.
\label{eq:m-three-body-adjoint-integral}\end{equation}

The angular integral retains \(|2\mathbf n-\mathbf r|>2\). The
first-order adjoint correction paired with the equilibrium radial flux
has zero angular average; the calculation is given in Supplementary
Section S1. Thus the zero-shear viscosity depends on this contact
projection of \(\psi_3\).

\section{3. The superposition
cancellation}\label{the-superposition-cancellation}

At first order in \(\mathrm{Pe}\), the superposition product becomes the
sum

\begin{equation}\psi_3^K=\psi_2(\mathbf r)+\psi_2(\mathbf s)+\psi_2(\mathbf s-\mathbf r).
\label{eq:m-superposition-response}\end{equation}

The superscript \(K\) denotes superposition. Inserting
\eqref{eq:m-superposition-response} into
\eqref{eq:m-three-body-adjoint-integral} separates three pair
contributions. Writing \(H_1=H(\mathbf r)\), \(H_2=H(2\mathbf n)\) and
\(H_3=H(2\mathbf n-\mathbf r)\) for the three terms, they are

\begin{equation}T_k=-4\int_{|\mathbf r|>2}d^3r\oint_{\mathrm{acc}}H_k(\mathbf r,2\mathbf n)\,\mathbf n\cdot\nabla H(\mathbf r)\,d\Omega_n,
\label{eq:m-pair-terms}\end{equation}

so that \(J_1^K=(64/9)(T_1+T_2+T_3)\). Table~\ref{tab:pair} gives their exact values.
With this normalization, \(c_3^K(0)=-(72/\pi^2)(T_1+T_2+T_3)\).

\begin{longtable}[]{@{}lr@{}}
\caption{Exact pair contributions to the zero-shear superposition
coefficient.\label{tab:pair}}\tabularnewline
\toprule\noalign{}
pair contribution & exact integral \\
\midrule\noalign{}
\endfirsthead
\toprule\noalign{}
pair contribution & exact integral \\
\midrule\noalign{}
\endhead
\bottomrule\noalign{}
\endlastfoot
first and second spheres, \(T_1\) & \(-9\pi^2/320\) \\
first and third spheres, \(T_2\) & \(-13\pi^2/960\) \\
second and third spheres, \(T_3\) & \(\pi^2/24\) \\
sum & \(0\) \\
\end{longtable}

The first term couples the shear distortion of the selected pair to the
equilibrium exclusion flux; the other two contain the shear distortions
of the two pairs that involve the third sphere. Each contribution is nonzero, but their sum
vanishes. Figure~\ref{fig:geometry}(b) shows the relevant overlap shell,
\(2<|\mathbf r|<4\), in which part of the contact sphere around sphere 1
is blocked by sphere 2. Contacts with sphere 1 are therefore missing on
one side, which gives the equilibrium exclusion flux; for
\(|\mathbf r|>4\) the excluded regions of spheres 1 and 2 separate and
this flux vanishes. Supplementary Section S1 evaluates the sum and specifies the
angular-before-radial integration order for the conditionally
convergent pieces.

The contact-density form of the same cancellation (Supplementary
Section S1.3) makes its physical content explicit. A third sphere near a
pair pushes the two spheres together, raising their equilibrium contact
value to \(1+5\phi/2\); this crowding amplifies the shear-induced excess
of compressional contacts and contributes \(+6\) to \(c_3\). The shear
distortions of the two pairs that involve the third sphere reduce the
contact distortion of the selected pair and contribute \(-3\) each. This
grouping by physical effect differs from the pair-by-pair grouping of
Table~\ref{tab:pair}, in which \(-(72/\pi^2)T_k\) gives \(81/40\),
\(39/40\) and \(-3\). The term \(T_3\) equals one of the \(-3\)
contributions (Supplementary Section S1.3), and \(T_1\) and \(T_2\)
together equal the remaining \(6-3=3\). Superposition contains both effects, the contact rise exactly
and the other-pair distortions in product form, and in it they cancel.
What it omits is the correction \(\chi\), the additional distortion
required by the shared contact constraints (Section 4).

This derivation recovers the cancellation identified in Lionberger's
thesis {[}10, Eqs. (A.61), (A.62){]}. The zero also follows from earlier
equations when they reduce to the same superposition sum {[}9, 14, 18, 19{]}. The reduction here adds the separate pair contributions, an
elementary radial identity for their sum and the relation of the zero to
the hard-sphere contact theorem, in the equal-sphere, zero-shear limit.
The exact cancellation within the closure and the accuracy of the
closure for the physical three-body response are separate questions.

For closures represented by weights \((w_1,w_2,w_3)\) in the three-term
functional,

\begin{equation}c_3(0)=-\frac{72}{\pi^2}(w_1T_1+w_2T_2+w_3T_3).
\label{eq:m-closure-weights}\end{equation}

Table~\ref{tab:closures} lists the resulting contact-stress coefficients and the sources
defining each closure. Hypernetted-chain and related closures share the
superposition row at this order. The weights describe reductions of the
stress functional; they do not imply a literal removal of a pair
interaction. The last two rows retain a separate equilibrium
contact-density contribution. Supplementary Section S1 gives the
reductions and stress comparisons.

\Needspace{23\baselineskip}
\begin{longtable}[]{@{}llr@{}}
\caption{Third-order coefficients of common closures, evaluated with the
contact stress.\label{tab:closures}}\tabularnewline
\toprule\noalign{}
closure in the dilute limit & retained terms & \(c_3(0)\) \\
\midrule\noalign{}
\endfirsthead
\toprule\noalign{}
closure in the dilute limit & retained terms & \(c_3(0)\) \\
\midrule\noalign{}
\endhead
\bottomrule\noalign{}
\endlastfoot
superposition and hypernetted-chain {[}3, 8, 9, 11{]} & \((1,1,1)\) &
\(0\) \\
equilibrium three-body distribution {[}20{]} & \((0,0,0)\) & \(0\) \\
Russel-Gast {[}12{]} & \((1,0,0)\) & \(81/40\) \\
S-PY and PYNEQ\textsuperscript{a} {[}13{]} & \((1,0,1)\) & \(-39/40\) \\
Percus-Yevick with the bare potential {[}9, 11{]} & \((0,1,1)\) &
\(-81/40\) \\
decoupled theory of Szamel {[}14{]} & contact rise\textsuperscript{b}
and \(T_3\) & \(3\) \\
Pair closure {[}3, 17{]} & contact rise\textsuperscript{b} & \(6\) \\
\end{longtable}

\textsuperscript{a} S-PY and PYNEQ are the Percus-Yevick-type
nonequilibrium closures of Wagner and Russel. \textsuperscript{b}
Contact rise is the separate \(+6\) equilibrium contact-density
contribution. The coefficients are our reductions in the equal-sphere,
zero-shear limit, using the contact stress; Lionberger's thesis gives the
same Russel-Gast value {[}10, p.~111, Eq. (4.111){]}.
Supplementary Section S1
gives source equations, distinguishes joint- and conditional-equilibrium
approximations, and derives the conversion between contact-density and
pair-distortion representations.

The physical three-body response must also satisfy the shared contact
constraints, whose error supplies the correction below.

\section{4. Contact correction and conditional viscosity
bounds}\label{contact-correction-and-conditional-viscosity-bounds}

\subsection{4.1 The boundary error and its remote
tail}\label{the-boundary-error-and-its-remote-tail}

The three-body perturbation is harmonic in the sphere positions. At
contact of spheres \(i\) and \(j\), with \(\mathbf n\) pointing from
\(i\) to \(j\), it obeys

\begin{equation}(\nabla_j-\nabla_i)\cdot\mathbf n\,\psi_3=2n_xn_y.
\label{eq:m-three-body-contact}\end{equation}

The perturbation of that pair supplies the required derivative. The
other two pair perturbations also vary when the contacting spheres move
apart. Their extra derivative produces the superposition flux error

\begin{equation}e=(\nabla_j-\nabla_i)\cdot\mathbf n\,\psi_3^K-2n_xn_y.
\label{eq:m-contact-flux-error}\end{equation}

For contact between spheres 1 and 2,
\(\nabla_2-\nabla_1=2\nabla_{\mathbf r}+\nabla_{\mathbf s}\). The two
other-pair derivatives have opposite signs, so at
\(\mathbf r=2\mathbf n\),

\begin{equation}e=\mathbf n\cdot[\nabla\psi_2(\mathbf s)-\nabla\psi_2(\mathbf s-\mathbf r)].
\label{eq:m-spectator-flux-difference}\end{equation}

Physically, \(e\) is a spurious probability flux through the contact of
spheres 1 and 2, the configuration of Fig.~\ref{fig:geometry}(c).
Superposition gives each pair its own distortion. When spheres 1 and 2
move apart along \(\mathbf n\) with sphere 3 held fixed, the distortions
of the pairs 1--3 and 2--3 change as well, and the two changes do not
cancel, so the product distribution
carries probability across a surface that the contact constraint makes impermeable.

When the third sphere is far away, Taylor expansion at fixed
\(\mathbf r\) gives

\begin{equation}e=2\mathbf n^{\mathsf T}\operatorname{Hess}\psi_2(\mathbf s)\mathbf n
+O(|\mathbf s|^{-6})=O(|\mathbf s|^{-5}).
\label{eq:m-spectator-source-tail}\end{equation}

Here \(\operatorname{Hess}\psi_2\) is the matrix of second derivatives.
Equation \eqref{eq:m-pair-response} has homogeneity \(-3\), its gradient
has homogeneity \(-4\), and the difference removes that leading
gradient. The source therefore has an integrable algebraic tail along a
remote spectator boundary. With the trace inequality of Section 4.2,
this decay makes the contact source a bounded functional on the energy
space; the conditions on the physical solution are stated there.

Write \(\psi_3=\psi_3^K+\chi\). The correction \(\chi\) is harmonic in
\(\Omega\) and cancels \(e\) on every contact surface. Its response to
this boundary source carries the contribution omitted by superposition.
The error \(e\) decays as \(|\mathbf s|^{-5}\) when the third sphere
leaves, so the source of the correction is concentrated on triplets in
which the third sphere is within a few radii of a contacting pair; these
are the configurations in which superposition misrepresents three-body
correlations.

\subsection{4.2 Conditional energy
identity}\label{conditional-energy-identity}

Let \(\mathcal D[f,h]\) be the bilinear form obtained by polarizing

\begin{equation}\mathcal D[f,f]=\iint_\Omega\left(|\nabla_{\mathbf r} f|^2+|\nabla_{\mathbf s} f|^2+
|\nabla_{\mathbf r} f+\nabla_{\mathbf s} f|^2\right)d^3r\,d^3s.
\label{eq:m-dirichlet-form}\end{equation}

Its integrand is the sum of the squared gradients with respect to the
three sphere positions, a positive quadratic form in the gradients. For a small distortion \(f\) of the uniform equilibrium distribution,
\(\mathcal D[f,f]\) is, at quadratic order in \(f\), proportional to the
rate at which free diffusion with impermeable contacts dissipates the
free energy stored in the distortion. Define \(\mathcal E[f]=\mathcal D[f,f]/6\) and the
one-contact pairing

\begin{equation}\langle f,e\rangle=\int d^3s\oint_{|\mathbf r|=2}f\,e\,dS_r,
\label{eq:m-contact-pairing}\end{equation}

restricted to accessible configurations. Sphere-label symmetry makes the
three contact contributions equal.

The energy space \(\mathcal H\) is the completion, in the norm
\(\mathcal D[f,f]^{1/2}\), of the restrictions to \(\Omega\) of smooth
compactly supported functions on the ambient six-dimensional space of
\((\mathbf r,\mathbf s)\) that are invariant under all six permutations
of the sphere labels. These functions need not vanish at contact, so the
elements of \(\mathcal H\) have finite energy and may have nonzero
contact traces. Let
\(\mathcal R^2=|\mathbf r|^2+|\mathbf s|^2+|\mathbf s-\mathbf r|^2\) be
the sum of the squared pair distances. Every \(f\in\mathcal H\)
satisfies the trace inequality

\begin{equation}72\int d^3s\oint_{|\mathbf r|=2}\frac{f^2}{\mathcal R^2}\,dS_r\le\mathcal D[f,f],
\label{eq:m-trace-inequality}\end{equation}

where the \(\mathbf s\) integral again covers accessible configurations.
The constant 72 is sharp: \(f=\mathcal R^{-4}\) belongs to
\(\mathcal H\) and gives equality. Supplementary Section S2.1 proves
\eqref{eq:m-trace-inequality} from a divergence identity for the
dilation field in center-of-mass coordinates, which is transverse to
every contact surface. The same identity gives
\(\iint_\Omega f^2\mathcal R^{-2}\,d^3r\,d^3s\le\mathcal D[f,f]/12\).
Hence \(\mathcal H\) is a space of functions, and
\eqref{eq:m-trace-inequality} defines their contact traces in a weighted
mean-square sense.

Physically, a distortion whose contact values are large in this
mean-square sense has a correspondingly large diffusive dissipation in
the surrounding volume; this is what makes the
contact source well defined.

The source tail and the trace inequality make \(\langle f,e\rangle\) a
bounded functional on \(\mathcal H\). The Riesz representation theorem
therefore gives a unique weak correction \(\chi_w\in\mathcal H\)
satisfying

\begin{equation}\mathcal D[f,\chi_w]=3\langle f,e\rangle\quad\text{for every }f\in\mathcal H.
\label{eq:m-weak-correction}\end{equation}

The factor 3 counts the three equal contact surfaces. For a smooth
harmonic correction whose contact derivative cancels \(e\), Green's
identity gives this same relation. Setting \(f=\chi_w\) yields
\(\mathcal E[\chi_w]=\langle\chi_w,e\rangle/2\). Its energy is finite
and positive, since the source pairing is bounded and does not vanish
identically. Supplementary Section S2 proves these statements. Two
conditions are needed to identify this weak solution with the physical
response and its viscosity.

\textbf{Proposition 1.} Assume that (i) the physical zero-shear
first-order correction \(\chi=\psi_3-\psi_3^K\) exists and equals
\(\chi_w\), and (ii) the remote remainder of the pair adjoint identity
\eqref{eq:m-adjoint-flux-identity} vanishes at first order in
\(\mathrm{Pe}\), as specified in Supplementary Section S1.2. Then

\begin{equation}c_3(0)=\frac{81}{32\pi^2}\mathcal E[\chi_w],\qquad
\mathcal E[\chi_w]=\frac12\langle\chi_w,e\rangle>0.
\label{eq:m-conditional-energy-identity}\end{equation}

\emph{Proof.} Condition (ii) gives the representation
\eqref{eq:m-three-body-adjoint-integral}. Its superposition part
vanishes by Table~\ref{tab:pair}. By condition (i) and sphere-label symmetry
(Supplementary Section S2.2), its correction part is
\(J_{1,\chi}=-\langle\chi_w,e\rangle/8\), a pairing that converges
absolutely by \eqref{eq:m-trace-inequality}. Combining this with
\eqref{eq:m-three-body-adjoint-integral} and
\(\mathcal E[\chi_w]=\langle\chi_w,e\rangle/2\) yields the coefficient
in \eqref{eq:m-conditional-energy-identity}, and the nonzero source
gives its positivity.

The linearized three-body equation, its contact conditions and the
source \(e\) are exact, and the two conditions of Proposition~1 concern only which solution of these equations is the physical one and whether a surface
term at infinity vanishes.
Condition (i) states that the physical first-order correction belongs to
the energy space \(\mathcal H\) and satisfies the weak equation
\eqref{eq:m-weak-correction}. Membership in \(\mathcal H\) means finite
diffusive dissipation together with the far-field behavior shared by the
elements of \(\mathcal H\) on every remote cluster end, that is,
wherever all three spheres separate or one sphere leaves a pair that
stays close. When one sphere recedes from a close pair, the pair should
recover its isolated shear response and the remaining three-body
correction should vanish. At pair level, the expected decay of
shear-induced correlations motivates condition (ii), which requires
the first-order outer surface integral in Supplementary Section S1.2 to
vanish at large separation. A vanishing far-field distortion is a
standard boundary condition for closed pair equations, as in the
calculations of Russel and Gast {[}12, Eq. (35){]} and Lionberger
and Russel {[}9, Eq. (51){]}. The adjoint field
decays by the explicit pair solution \eqref{eq:m-pair-response}, while
the physical pair-density correction depends on the three-body response.
The energy estimate controls
the weak correction; identifying it with the physical response and showing
that the remote remainder vanishes require separate arguments. We therefore state both conditions as assumptions.
Supplementary Sections S1 and S2 state the conditions
separately from the admissibility of an explicit trial.

Define the weak-energy coefficient

\begin{equation}\Theta=\frac{81}{32\pi^2}\mathcal E[\chi_w].
\label{eq:m-weak-coefficient}\end{equation}

The bounds below hold for \(\Theta\) unconditionally; Proposition~1
identifies \(\Theta\) with \(c_3(0)\).

\subsection{4.3 Variational lower bound}\label{variational-lower-bound}

For every \(f\in\mathcal H\), \eqref{eq:m-weak-correction} gives
\(\mathcal D[f,\chi_w]/6=\langle f,e\rangle/2\). Completing the square
yields

\begin{equation}\mathcal E[\chi_w]=\max_{f\in\mathcal H}\left(\langle f,e\rangle-\frac16\mathcal D[f,f]\right).
\label{eq:m-dirichlet-principle}\end{equation}

Consequently, every admissible \(f\), that is every \(f\in\mathcal H\),
gives

\begin{equation}F(f)=\frac{81}{32\pi^2}\left(\langle f,e\rangle-\frac16\mathcal D[f,f]\right)\le\Theta.
\label{eq:m-trial-functional}\end{equation}

This is a Dirichlet variational principle. Related principles bound
sedimentation {[}21{]}, intrinsic viscosity {[}22{]} and the shear
viscosity of crosslinked melts with excluded-volume interactions
{[}23{]}. Here the source is the contact error of superposition.

The lower trial is a linear combination of 500 symmetric basis
functions, formed from directional derivatives of the pair response and
summed over all sphere relabelings. Their decay on every remote cluster
end ensures membership in \(\mathcal H\). Supplementary Section S2 gives
the complete basis, coefficients and integration formulas.

Rotation averaging and scale integration reduce the entries to
\(\pi^2(x_0+x_1\ln2)+\pi^4x_2\), with rational \(x_0,x_1,x_2\). Exact
integration followed by outward rational arithmetic evaluates the
functional of the specified trial. The resulting lower bound is

\begin{equation}\Theta\ge1.8266948.
\label{eq:m-numerical-lower-bound}\end{equation}

This conclusion uses admissibility and the evaluated functional; it does
not require the trial to maximize over its finite space. Under
Proposition 1 it is also a lower bound on \(c_3(0)\).

\subsection{4.4 A conditional two-sided
bound}\label{a-conditional-two-sided-bound}

The lower functional measures the energy captured by a trial. Its gap
from the weak-energy coefficient is exactly
\(K\mathcal D[\chi_w-v,\chi_w-v]/6\), where \(K=81/(32\pi^2)\).
Integration by parts in the weak error equation expresses this unknown
error through two computable failures of \(v\): its bulk residual
\(\mathcal Lv\), which measures departure from the harmonic equation,
and its contact residual \(b_v\), which measures the remaining flux
error. The dual norm of this residual functional bounds the error
energy. The radial and rotational energy retained below supplies a
computable upper bound on that norm. The lower bound thus requires
finite energy, whereas this upper estimate additionally requires finite
residual norms.

Rotating the three-sphere configuration changes its shear response with
the quadrupolar symmetry of the imposed strain, the \(l=2\) rotation
sector. This symmetry supplies an additional positive term in the energy
estimate. Retaining that term and the radial gradient energy gives a
sharper estimate of the residual than separate volume and contact
inequalities.

Let
\(\mathcal L=2\Delta_{\mathbf r}+2\Delta_{\mathbf s}+2\nabla_{\mathbf r}\cdot\nabla_{\mathbf s}\)
and \(b_v=e+\mathbf n\cdot(2\nabla_{\mathbf r}+\nabla_{\mathbf s})v\) on
\(|\mathbf r|=2\). In addition to finite energy, the upper trial has
finite weighted volume and contact norms

\begin{equation}A_v^2=\iint_\Omega\mathcal R^2(\mathcal Lv)^2\,d^3r\,d^3s,
\qquad B_v^2=\int d^3s\oint_{|\mathbf r|=2}\mathcal R^2b_v^2\,dS_r.
\label{eq:m-upper-residual-norms}\end{equation}

Finiteness of these norms makes the volume and contact source pairings
continuous on the energy space. The bound below retains their signed
coupling.

At fixed triangle shape, write the shortest pair distance as
\(\ell=2e^x\). Rescaling a test field as \(f=\ell^{-2}w(x)\) converts
the retained radial and rotational energy to a half-line quadratic form
with operator \(-\partial_x^2+10\) and homogeneous Robin condition
\(w'(0)=2w(0)\). The contact residual acts as a boundary source. The
Green function of this operator evaluates a computable squared residual
norm \(\eta(v)\), including the signed interaction between the volume
and contact sources. Dropping the remaining nonnegative shape-gradient
energy enlarges the test space, so

\begin{equation}\Theta\le F(v)+\frac{27}{64\pi^2}\eta(v).
\label{eq:m-residual-upper-bound}\end{equation}

Physically, the upper bound is the energy captured by the trial plus a
bound on the energy that the trial misses, computed from its two
remaining errors:
departure from free diffusion in the bulk and leakage through the
contacts.

Supplementary Section S2 derives the Green function, normalization and
exact integration of \(\eta(v)\). A separate 496-mode trial satisfies
the weighted residual conditions and gives the upper endpoint below. Its
complete coefficients are listed alongside those of the lower trial.
Both fixed-trial evaluations use exact symbolic integrals and outward
rational bounds, with independent evaluations of the complete functions.

\begin{equation}1.8266948\le\Theta\le1.8267157,
\qquad c_3(0)=\Theta\quad\text{under Proposition 1}.
\label{eq:m-two-sided-bound}\end{equation}

The interval has width \(0.0000209\), less than \(0.00115\) percent of
its lower endpoint. All the closure coefficients in
Table~\ref{tab:closures} lie outside it, so under Proposition~1 each of
them differs from \(c_3(0)\).

\section{5. Discussion}\label{discussion}

The contact constraints supply the physical content that pair-product
closures omit, and they connect the low-density closure benchmark of
Lionberger and Russel to the full three-body response.

\subsection{5.1 What is proved and what is conditional}\label{proved-and-conditional}

Three statements are proved without invoking Proposition 1. Kirkwood
superposition gives exactly zero for the contact-stress coefficient
(Table~\ref{tab:pair}); the contact-density reduction of Supplementary
Section S1.3 obtains the same zero from an integral over a bounded
region, independently of both conditions of Proposition 1. The superposition flux error on the shared contacts defines a unique
finite-energy weak correction with positive dissipation. The weak-energy coefficient \(\Theta\) is a well-defined property of
the exact linearized three-body problem, and two explicit trials
evaluated with outward rational arithmetic give
\(1.8266948\le\Theta\le1.8267157\). The interval for the physical
coefficient is conditional: it applies to \(c_3(0)\) when the two
conditions of Proposition 1 hold. Proving the conditions requires far-field
estimates for the physical three-body perturbation and for the first density correction beyond those that the energy
method supplies.

The two functionals also apply to fields other than the trials used here: a closure that
supplies an admissible correction can be assessed through
\eqref{eq:m-trial-functional}, and a trial with finite residual norms
supplies an upper bound through \eqref{eq:m-residual-upper-bound}.

\subsection{5.2 Physical mechanism of the positive correction}\label{mechanism}

The Brownian viscosity measures the rate at which Brownian motion must
dissipate the distortion that the flow imposes on the microstructure. At
the pair level, we insert the pair response \eqref{eq:m-pair-response}
into the contact stress. Integration by parts, using
\(\nabla^2\psi_2=0\) and the contact condition, gives

\begin{equation}\frac{12}{5}=\frac{27}{8\pi}\int_{|\mathbf r|>2}|\nabla\psi_2|^2\,d^3r,
\label{eq:m-pair-dissipation}\end{equation}

the diffusive dissipation of the pair distortion.
Equation~\eqref{eq:m-conditional-energy-identity} is the three-body
analog. The three pair distortions in superposition make no net
contribution to the contact stress. The distortions of the two pairs
involving the third sphere cancel the crowding-induced rise of the
contact value (Section 3). What superposition omits is the correction \(\chi\), the additional
distortion required to stop the spurious flux \(e\) through every
contact, Fig.~\ref{fig:geometry}(c).
Enforcing all three contact constraints at once requires this
additional distortion. Under Proposition 1, its positive dissipation
gives a positive \(c_3(0)\) and therefore raises the Brownian viscosity at
third order. The source \(e\) decays as \(|\mathbf s|^{-5}\), so this
distortion is generated by triplets near contact.

\subsection{5.3 Relation to the classical viscosity expansion and to experiments}\label{classical}

In real suspensions hydrodynamic interactions contribute at every order
beyond the Einstein term. For near-hard spheres with full hydrodynamic
interactions at zero shear, Bergenholtz et al. evaluate
\(\eta/\eta_s=1+(5/2)\phi+5.92\phi^2\), of which \(5.00\phi^2\) is the
hydrodynamic high-frequency term and only \(0.92\phi^2\) is Brownian,
against \(12/5\) without hydrodynamic interactions
{[}2, pp.~252--253{]}. At third order, the accurately computed
coefficient is the high-frequency value \(9.09\pm0.02\) for hard spheres
at equilibrium structure {[}24{]}; it includes pair hydrodynamics
weighted by the density correction to the pair distribution as well as
three-particle hydrodynamic interactions. The coefficient \(c_3(0)\) is
the Brownian structural term of the model without hydrodynamic
interactions. By contrast, \(9.09\) is a hydrodynamic coefficient, and a
total-viscosity coefficient combines both kinds of stress. Hydrodynamic
interactions add their own third-order stress and also change the Brownian contribution itself, as
the fall from \(12/5\) to \(0.92\) at second order shows. Closure
estimates that keep only pairwise-additive hydrodynamic interactions
exist {[}13, Table V{]}. The third-order Brownian coefficient with
full hydrodynamic interactions has, to our knowledge, not been evaluated: it
requires the solution of the three-particle Smoluchowski equation, which
Wajnryb and Dahler formulated but did not solve {[}25, p.~257{]}.

Rheometry measures total viscosities.
In noncolloidal suspensions the Brownian stress is negligible. Pasquino et al.
measured poly(methyl methacrylate) spheres under conditions in which
Brownian motion and inertia are negligible. In a Newtonian matrix the
quadratic coefficient of the shear viscosity was \(7.00\pm0.18\), with
the linear coefficient fixed at \(5/2\) {[}26, Table II{]}. This value lies within the
error bound of Batchelor and Green's estimate \(7.6\pm0.8\), a purely
hydrodynamic result for steady pure straining. In simple shear their
theory gives no value: hydrodynamics alone leaves the pair
distribution indeterminate on closed trajectories
{[}6, pp.~401--402{]}. Colloidal hard-sphere dispersions contain
the Brownian stress, but always together with hydrodynamic coupling
{[}27--29{]}, so relating their viscosity to
Eq.~\eqref{eq:m-viscosity-expansion} requires a matched dynamical model
and stress definition. Stress fluctuations imaged in quiescent, nearly
hard-sphere dispersions give the Brownian viscosity through the
Green-Kubo relation at volume fractions from 0.12 to 0.45
{[}30{]}, again with hydrodynamic coupling. At third order, de Kruif et al. fitted a cubic
coefficient of \(42\pm10\) to the zero-shear viscosity of sterically
stabilized silica spheres {[}27{]}; its uncertainty alone is more than
five times \(\Theta\approx1.83\).

\subsection{5.4 Role of the coefficient and its measurement}\label{role}

The coefficient \(c_3(0)\) provides a benchmark for this model.
Brownian dynamics simulations without hydrodynamic interactions can
determine it, and closure theories formulated in the same model
approximate it. Closures of the kind compared in Table~\ref{tab:closures}
also supply the three-body term of microstructural theories that include
hydrodynamic interactions {[}19{]}. In the model without hydrodynamic
interactions, the exact three-body equations provide a reference for
this term. Under Proposition~1, \(c_3(0)\) lies in the interval
\eqref{eq:m-two-sided-bound}, which excludes every closure coefficient
compared here. The shared contact constraints generate a positive
correction. Under Proposition~1, this correction is the whole of
\(c_3(0)\).

The effect of the coefficient on the stress can be illustrated by
truncating the viscosity expansion after the cubic term. At volume fraction \(\phi=0.1\) the pair contribution is \(0.024\)
and, under Proposition~1, the cubic contribution is \(0.0018267\), so
the cubic truncation is \(0.025827\), compared with \(0.024\) for
Kirkwood superposition and \(0.026025\) for the Russel-Gast coefficient
\(81/40\). These values concern the Brownian viscosity normalized by the viscosity
of the suspending fluid; a total-viscosity prediction also requires the relevant
hydrodynamic response.

The same numbers set the precision that a simulation test needs. At
\(\phi=0.05\), once the higher-order contributions are separated, telling
\(c_3(0)\) apart from the nearest closure coefficient, \(81/40\), by two
standard deviations requires a standard deviation of about 0.2\% in the
Brownian viscosity. A test therefore subtracts the known pair term at
several low volume fractions and separates the cubic contribution from
the higher-order ones.

Existing finite-concentration simulations offer a consistency comparison
for the sum of the cubic and higher-order terms. Lowe and Masters report a Brownian viscosity
\(0.026\pm0.001\) at \(\phi=0.1\), with the quoted uncertainty equal to
two standard deviations {[}4{]}. Subtracting the pair term gives
\(c_3+0.1c_4+\cdots=2.0\pm0.5\) at one statistical standard deviation;
rounding of the quoted 0.026 contributes an uncertainty of comparable
size. After the pair term is subtracted, the viscosities that Foss and
Brady obtained by Green-Kubo Brownian dynamics {[}5, Table I{]} give
effective residuals 4.4 at \(\phi=0.2\) and 7.3 at \(\phi=0.3\), which
combine cubic and higher-order terms. The viscosities were computed at a
fixed time step of \(2.5\times10^{-4}a^2/D_0\), without time-step
extrapolation or quoted uncertainties. Swaroop's Green-Kubo simulations of
the same model, also at a fixed time step, report
\(0.0060\pm0.0002\) at \(\phi=0.05\) and \(0.0283\pm0.0005\) at
\(\phi=0.1\) {[}31, Table 5.12{]}; with the quoted errors the
residuals are \(0.0\pm1.6\) and \(4.3\pm0.5\). At \(\phi=0.1\) this
viscosity and that of Lowe and Masters differ by 0.0023, more than the
cubic contribution 0.0018, so the existing simulations do not yet
resolve the cubic term.

For an excluded radius \(b\) and hydrodynamic radius \(a\), dimensional
rescaling of the same contact-diffusion model gives

\begin{equation}\frac{\eta_B}{\eta_s}=\frac{12}{5}(b/a)^5\phi^2+c_3(0)(b/a)^8\phi^3+\cdots,
\label{eq:m-excluded-radius-scaling}\end{equation}

with \(\phi\) based on the hydrodynamic spheres; the pair term of this
form is given in {[}1{]}. The strong radius dependence makes the
three-body contribution relevant when exclusion occurs at separations
where hydrodynamic coupling is weak. Application to a suspension with
appreciable hydrodynamic coupling requires its mobility-dependent pair
and three-body response.

\section{6. Conclusions}\label{conclusions}

Kirkwood superposition gives exactly zero for the third-order
coefficient of the zero-shear Brownian viscosity of hard spheres without
hydrodynamic interactions: three nonzero pair contributions
cancel. Its error is a flux through the shared contacts, and that error defines
a unique weak correction with positive dissipation. Two explicit trials
prove that the corresponding weak-energy coefficient lies between
1.8266948 and 1.8267157. For the physical coefficient \(c_3(0)\) this is a conditional interval:
it applies when the physical three-body correction equals the weak
correction and the remote remainder of the pair adjoint identity
vanishes, the two conditions of Proposition 1. As a benchmark for the model without hydrodynamic
interactions, the interval excludes every closure coefficient compared here, and under these conditions the positive
correction generated by the shared contact constraints carries the whole
coefficient.

\section{Declarations}\label{declarations}

\subsection{Acknowledgments}\label{acknowledgments}

The authors gratefully acknowledge the Advanced Cyberinfrastructure for
Education and Research (ACER) group at the University of Illinois at
Chicago for providing computational resources and services needed to
deliver research results delivered within this paper. URL:
https://acer.uic.edu. Computations used the Lakeshore high-performance
computing cluster.

Codex (OpenAI) assisted in developing and verifying the variational
lower bound and the shear-sector refinement and certification of the
upper bound. Claude (Anthropic) assisted in developing and verifying the
Hardy-type trace estimate and residual upper-bound formulation. Both
tools also assisted with derivations, implementation, literature checks,
manuscript drafting and revision. The authors are responsible for the
accuracy, originality and final content of this work.

\subsection{Research ethics}\label{research-ethics}

Not applicable.

\subsection{Informed consent}\label{informed-consent}

Not applicable.

\subsection{Author contributions}\label{author-contributions}

All authors have accepted responsibility for the entire content of this
manuscript and approved its submission.

\textbf{Julian Olszewski:} Conceptualization (equal); Funding
acquisition (lead); Investigation (equal); Methodology (equal); Writing
-- original draft (equal); Writing -- review \& editing (equal);
Software; Formal analysis; Data curation; Visualization.

\textbf{Zijian Liu:} Conceptualization (equal); Investigation (equal);
Methodology (equal); Writing -- original draft (equal); Writing --
review \& editing (equal); Software; Formal analysis; Data curation;
Visualization.

\textbf{Jie Xu:} Conceptualization (equal); Supervision (lead); Writing
-- review \& editing (equal).

\subsection{Conflict of interest}\label{conflict-of-interest}

The authors state no conflict of interest.

\subsection{Research funding}\label{research-funding}

J. Olszewski gratefully acknowledges support from the Chancellor's
Undergraduate Research Award (CURA), University of Illinois Chicago.

\subsection{Data availability}\label{data-availability}

The authors confirm that the data supporting the findings of this study
are available within the article and its supplementary materials.

\begingroup\setlength{\parskip}{3pt plus 1pt}
\section{References}\label{references}

\noindent\begin{minipage}{\linewidth}
{[}1{]} B. Cichocki and B. U. Felderhof, ``Linear viscoelasticity of semidilute hard-sphere suspensions,'' \emph{Phys. Rev.~A}, vol.~43, no. 10, pp.~5405--5411, 1991 (doi: 10.1103/PhysRevA.43.5405).
\end{minipage}\par

\noindent\begin{minipage}{\linewidth}
{[}2{]} J. Bergenholtz, J. F. Brady, and M. Vicic, ``The non-Newtonian rheology of dilute colloidal suspensions,'' \emph{J. Fluid Mech.}, vol.~456, pp.~239--275, 2002 (doi: 10.1017/S0022112001007583).
\end{minipage}\par

\noindent\begin{minipage}{\linewidth}
{[}3{]} R. A. Lionberger and W. B. Russel, ``Microscopic theories of the rheology of stable colloidal dispersions,'' in \emph{Advances in Chemical Physics}, vol.~111, I. Prigogine and S. A. Rice, Eds., New York, NY, USA, John Wiley \& Sons, 2000, pp.~399--474 (ISBN: 0-471-34990-9; doi: 10.1002/9780470141700.ch3).
\end{minipage}\par

\noindent\begin{minipage}{\linewidth}
{[}4{]} C. P. Lowe and A. J. Masters, ``The viscoelastic response of Brownian suspensions,'' \emph{J. Chem. Phys.}, vol.~111, no. 18, pp.~8708--8720, 1999 (doi: 10.1063/1.480211).
\end{minipage}\par

\noindent\begin{minipage}{\linewidth}
{[}5{]} D. R. Foss and J. F. Brady, ``Brownian dynamics simulation of hard-sphere colloidal dispersions,'' \emph{J. Rheol.}, vol.~44, no. 3, pp.~629--651, 2000 (doi: 10.1122/1.551104).
\end{minipage}\par

\noindent\begin{minipage}{\linewidth}
{[}6{]} G. K. Batchelor and J. T. Green, ``The determination of the bulk stress in a suspension of spherical particles to order \(c^2\),'' \emph{J. Fluid Mech.}, vol.~56, no. 3, pp.~401--427, 1972 (doi: 10.1017/S0022112072002435).
\end{minipage}\par

\noindent\begin{minipage}{\linewidth}
{[}7{]} G. K. Batchelor, ``The effect of Brownian motion on the bulk stress in a suspension of spherical particles,'' \emph{J. Fluid Mech.}, vol.~83, no. 1, pp.~97--117, 1977 (doi: 10.1017/S0022112077001062).
\end{minipage}\par

\noindent\begin{minipage}{\linewidth}
{[}8{]} J. G. Kirkwood, ``Statistical mechanics of fluid mixtures,'' \emph{J. Chem. Phys.}, vol.~3, no. 5, pp.~300--313, 1935 (doi: 10.1063/1.1749657).
\end{minipage}\par

\noindent\begin{minipage}{\linewidth}
{[}9{]} R. A. Lionberger and W. B. Russel, ``Effectiveness of nonequilibrium closures for the many body forces in concentrated colloidal dispersions,'' \emph{J. Chem. Phys.}, vol.~106, no. 1, pp.~402--416, 1997 (doi: 10.1063/1.473029).
\end{minipage}\par

\noindent\begin{minipage}{\linewidth}
{[}10{]} R. A. Lionberger, ``Rheology, structure, and diffusion in concentrated colloidal dispersions,'' Ph.D.~dissertation, Dept. Chem. Eng., Princeton Univ., Princeton, NJ, USA, 1996.
\end{minipage}\par

\noindent\begin{minipage}{\linewidth}
{[}11{]} K. Scherwinski, ``Two approximations for the classical triplet distribution function,'' \emph{Mol. Phys.}, vol.~70, no. 5, pp.~797--809, 1990 (doi: 10.1080/00268979000101361).
\end{minipage}\par

\noindent\begin{minipage}{\linewidth}
{[}12{]} W. B. Russel and A. P. Gast, ``Nonequilibrium statistical mechanics of concentrated colloidal dispersions: Hard spheres in weak flows,'' \emph{J. Chem. Phys.}, vol.~84, no. 3, pp.~1815--1826, 1986 (doi: 10.1063/1.450428).
\end{minipage}\par

\noindent\begin{minipage}{\linewidth}
{[}13{]} N. J. Wagner and W. B. Russel, ``Nonequilibrium statistical mechanics of concentrated colloidal dispersions: Hard spheres in weak flows with many-body thermodynamic interactions,'' \emph{Physica A}, vol.~155, no. 3, pp.~475--518, 1989 (doi: 10.1016/0378-4371(89)90003-4).
\end{minipage}\par

\noindent\begin{minipage}{\linewidth}
{[}14{]} G. Szamel, ``Nonequilibrium structure and rheology of concentrated colloidal suspensions: Linear response,'' \emph{J. Chem. Phys.}, vol.~114, no. 19, pp.~8708--8717, 2001 (doi: 10.1063/1.1366335).
\end{minipage}\par

\noindent\begin{minipage}{\linewidth}
{[}15{]} B. U. Felderhof and R. B. Jones, ``Linear response theory of the viscosity of suspensions of spherical Brownian particles,'' \emph{Physica A}, vol.~146, no. 3, pp.~417--432, 1987 (doi: 10.1016/0378-4371(87)90278-0).
\end{minipage}\par

\noindent\begin{minipage}{\linewidth}
{[}16{]} J. Bławzdziewicz and G. Szamel, ``Structure and rheology of semidilute suspension under shear,'' \emph{Phys. Rev.~E}, vol.~48, no. 6, pp.~4632--4636, 1993 (doi: 10.1103/PhysRevE.48.4632).
\end{minipage}\par

\noindent\begin{minipage}{\linewidth}
{[}17{]} J. F. Brady, ``The rheological behavior of concentrated colloidal dispersions,'' \emph{J. Chem. Phys.}, vol.~99, no. 1, pp.~567--581, 1993 (doi: 10.1063/1.465782).
\end{minipage}\par

\noindent\begin{minipage}{\linewidth}
{[}18{]} T. Ohtsuki, ``Dynamical properties of strongly interacting Brownian particles. I. Dynamic shear viscosity,'' \emph{Physica A}, vol.~108, nos. 2--3, pp.~441--458, 1981 (doi: 10.1016/0378-4371(81)90141-2).
\end{minipage}\par

\noindent\begin{minipage}{\linewidth}
{[}19{]} E. Nazockdast and J. F. Morris, ``Microstructural theory and the rheology of concentrated colloidal suspensions,'' \emph{J. Fluid Mech.}, vol.~713, pp.~420--452, 2012 (doi: 10.1017/jfm.2012.467).
\end{minipage}\par

\noindent\begin{minipage}{\linewidth}
{[}20{]} S. M. Tschopp and J. M. Brader, ``Superadiabatic dynamical density functional theory for colloidal suspensions under homogeneous steady-shear,'' \emph{J. Chem. Phys.}, vol.~160, no. 21, 2024, Art. no. 214124 (doi: 10.1063/5.0211198).
\end{minipage}\par

\noindent\begin{minipage}{\linewidth}
{[}21{]} J. Rotne and S. Prager, ``Variational treatment of hydrodynamic interaction in polymers,'' \emph{J. Chem. Phys.}, vol.~50, no. 11, pp.~4831--4837, 1969 (doi: 10.1063/1.1670977).
\end{minipage}\par

\noindent\begin{minipage}{\linewidth}
{[}22{]} M. Fixman, ``Variational bounds for polymer transport coefficients,'' \emph{J. Chem. Phys.}, vol.~78, no. 3, pp.~1588--1593, 1983 (doi: 10.1063/1.444849).
\end{minipage}\par

\noindent\begin{minipage}{\linewidth}
{[}23{]} C. H. Köhler, H. Löwe, P. Müller, and A. Zippelius, ``Variational bounds for the shear viscosity of gelling melts,'' \emph{EPL}, vol.~78, no. 4, 2007, Art. no. 46002 (doi: 10.1209/0295-5075/78/46002).
\end{minipage}\par

\noindent\begin{minipage}{\linewidth}
{[}24{]} B. Cichocki, M. L. Ekiel-Jeżewska, and E. Wajnryb, ``Three-particle contribution to effective viscosity of hard-sphere suspensions,'' \emph{J. Chem. Phys.}, vol.~119, no. 1, pp.~606--619, 2003 (doi: 10.1063/1.1576378).
\end{minipage}\par

\noindent\begin{minipage}{\linewidth}
{[}25{]} E. Wajnryb and J. S. Dahler, ``The Newtonian viscosity of a moderately dense suspension,'' in \emph{Advances in Chemical Physics}, vol.~102, I. Prigogine and S. A. Rice, Eds., New York, NY, USA, John Wiley \& Sons, 1997, pp.~193--313 (ISBN: 0-471-19144-2; doi: 10.1002/9780470141618.ch4).
\end{minipage}\par

\noindent\begin{minipage}{\linewidth}
{[}26{]} R. Pasquino, N. Grizzuti, P. L. Maffettone, and F. Greco, ``Rheology of dilute and semidilute noncolloidal hard sphere suspensions,'' \emph{J. Rheol.}, vol.~52, no. 6, pp.~1369--1384, 2008 (doi: 10.1122/1.2992600).
\end{minipage}\par

\noindent\begin{minipage}{\linewidth}
{[}27{]} C. G. de Kruif, E. M. F. van Iersel, A. Vrij, and W. B. Russel, ``Hard sphere colloidal dispersions: Viscosity as a function of shear rate and volume fraction,'' \emph{J. Chem. Phys.}, vol.~83, no. 9, pp.~4717--4725, 1985 (doi: 10.1063/1.448997).
\end{minipage}\par

\noindent\begin{minipage}{\linewidth}
{[}28{]} J. C. van der Werff, C. G. de Kruif, C. Blom, and J. Mellema, ``Linear viscoelastic behavior of dense hard-sphere dispersions,'' \emph{Phys. Rev.~A}, vol.~39, no. 2, pp.~795--807, 1989 (doi: 10.1103/PhysRevA.39.795).
\end{minipage}\par

\noindent\begin{minipage}{\linewidth}
{[}29{]} Z. Cheng, J. Zhu, P. M. Chaikin, S.-E. Phan, and W. B. Russel, ``Nature of the divergence in low shear viscosity of colloidal hard-sphere dispersions,'' \emph{Phys. Rev.~E}, vol.~65, no. 4, 2002, Art. no. 041405 (doi: 10.1103/PhysRevE.65.041405).
\end{minipage}\par

\noindent\begin{minipage}{\linewidth}
{[}30{]} N. Y. C. Lin, M. Bierbaum, and I. Cohen, ``Determining quiescent colloidal suspension viscosities using the Green-Kubo relation and image-based stress measurements,'' \emph{Phys. Rev. Lett.}, vol.~119, no. 13, 2017, Art. no. 138001 (doi: 10.1103/PhysRevLett.119.138001).
\end{minipage}\par

\noindent\begin{minipage}{\linewidth}
{[}31{]} M. Swaroop, ``The bulk viscosity of suspensions,'' Ph.D.~dissertation, California Inst. Technol., Pasadena, CA, USA, 2010 (doi: 10.7907/HQGZ-DV22).
\end{minipage}\par

\endgroup
\clearpage
\setcounter{equation}{0}
\renewcommand{\theequation}{S\arabic{equation}}
\setcounter{figure}{0}
\setcounter{table}{0}
\renewcommand{\thefigure}{S\arabic{figure}}
\renewcommand{\thetable}{S\arabic{table}}
\begin{center}
{\Large Supplementary Material}\\[6pt]
Three-body corrections to the Brownian viscosity of hard spheres without hydrodynamic interactions\\[6pt]
Julian Olszewski, Zijian Liu and Jie Xu
\end{center}

Section S1 derives the contact-stress normalization, the pair adjoint
identity and the closure reductions. Section S2 constructs the weak
correction and specifies the two global trial fields and their exact
interval evaluation. The enclosure applies to the weak coefficient
without a physical-response assumption; its interpretation as the
zero-shear viscosity coefficient uses the two conditions in Proposition
1 of the main article.

\section{S1. Contact stress, adjoint identity and closure
reductions}\label{s1.-contact-stress-adjoint-identity-and-closure-reductions}

\subsection{S1.1 Conventions and stress
normalization}\label{s1.1-conventions-and-stress-normalization}

Lengths and times use the sphere radius \(a\) and \(a^2/D_0\), with
\(D_0\) the free-sphere diffusivity. We use the relative coordinates of
Section 2.1 of the main article, with \(\mathbf q=\mathbf s-\mathbf r\)
and \(\Omega=\{r>2,s>2,q>2\}\). Here \(\mathrm{Pe}=\dot\gamma a^2/D_0\),
\(\dot\gamma\) is the shear rate, \(\eta_s\) is the viscosity of the
suspending fluid and \(\phi\) is the volume fraction. The dimensional
Brownian contact stress without hydrodynamic interactions is {[}1, Eq. (3.24),
with the hydrodynamic functions removed{]}, {[}2, Eq. (9){]}

\begin{equation}
\sigma^B_{xy}=-\frac{9k_BT\phi^2}{4\pi^2a^3}
\oint n_xn_y\,g(2a\mathbf n;\phi)\,d\Omega_n.
\label{eq:s-contact-stress}
\end{equation}

Here \(k_B\) is Boltzmann's constant and \(T\) is temperature. The
Stokes--Einstein relation \(D_0=k_BT/(6\pi\eta_s a)\) gives

\begin{equation}
\frac{9k_BT\phi^2}{4\pi^2a^3\eta_s\dot\gamma}
=\frac{27\phi^2D_0}{2\pi a^2\dot\gamma}
=\frac{27\phi^2}{2\pi\mathrm{Pe}}.
\label{eq:s-stress-normalization}
\end{equation}

Thus the contact argument is \(2\mathbf n\) in dimensionless
coordinates. A diameter-based Péclet number is four times the present
one. All closure values below use Eq.~\eqref{eq:s-contact-stress}; a
value obtained with a mean-force stress is identified separately.

At first order in shear, \(g_2=1+\mathrm{Pe}\,\psi_2+O(\mathrm{Pe}^2)\),
with \(\psi_2(\mathbf r)=-(16/3)H(\mathbf r)\) and
\(H(\mathbf r)=r_xr_y/r^5\). The superposition product {[}3, Eq. (40){]}
gives \(\psi_3^K=\psi_2(\mathbf r)+\psi_2(\mathbf s)+\psi_2(\mathbf q)\).

\subsection{S1.2 Pair hierarchy and adjoint
reduction}\label{s1.2-pair-hierarchy-and-adjoint-reduction}

Expand the pair distribution in dimensionless number density as
\(g=g_2+\rho g^{(1)}+O(\rho^2)\), with \(\rho=3\phi/(4\pi)\). The
three-body distribution entering this first density correction is
evaluated at infinite dilution and the same shear rate. In the pair
exterior \(\mathcal P=\{r>2\}\), put \(\mathbf u=r_y\mathbf e_x\),
\(\mathbf n_r=\mathbf r/r\) and define

\begin{equation}
\boldsymbol Q(\mathbf r)=-4\oint_{|2\mathbf n-\mathbf r|>2}
g_3(\mathbf r,2\mathbf n)\,\mathbf n\,d\Omega_n.
\label{eq:s-contact-flux}
\end{equation}

The factor four converts \(d\Omega_n\) to contact area. Integration over
the third sphere gives

\begin{equation}
\begin{aligned}
\mathcal A&=2\nabla_r^2-\mathrm{Pe}\,\mathbf u\cdot\nabla_r,\\
\mathcal A g^{(1)}&=-2\nabla_r\cdot\boldsymbol Q, &&r>2,\\
2\partial_r g^{(1)}
&=\mathrm{Pe}(\mathbf u\cdot\mathbf n_r)g^{(1)}
-2\mathbf n_r\cdot\boldsymbol Q, &&r=2,\\
g^{(1)}&\longrightarrow0, &&r\longrightarrow\infty.
\end{aligned}
\label{eq:s-pair-hierarchy}
\end{equation}

Equivalently,
\(\mathbf j^{(1)}=\mathrm{Pe}\,\mathbf u g^{(1)}-2\nabla_r g^{(1)}-2\boldsymbol Q\)
is divergence free and has zero normal contact flux. For the stress
projection \(m(\mathbf n_r)=n_{r,x}n_{r,y}\) choose

\begin{equation}
\begin{aligned}
\mathcal A^*&=2\nabla_r^2+\mathrm{Pe}\,\mathbf u\cdot\nabla_r,\\
\mathcal A^*w&=0, &&r>2,\\
2\partial_rw&=m/4, &&r=2,\\
w&\longrightarrow0, &&r\longrightarrow\infty.
\end{aligned}
\label{eq:s-pair-adjoint}
\end{equation}

The reversed shear follows from \(\nabla_r\cdot\mathbf u=0\); the factor
\(1/4\) converts solid angle to area. Green's identity on
\(\mathcal P_R=\{2<r<R\}\) gives

\begin{equation}
\begin{aligned}
J_{xy}&=\oint m\,g^{(1)}(2\mathbf n_r)\,d\Omega_r\\
&=2\int_{\mathcal P_R}\boldsymbol Q\cdot\nabla_rw\,d^3r-\mathcal B_R,\\
\mathcal B_R&=\oint_{r=R}\Big[2(w\partial_rg^{(1)}-g^{(1)}\partial_rw)\\
&\hspace{13mm}-\mathrm{Pe}(\mathbf u\cdot\mathbf n_r)wg^{(1)}
+2w\boldsymbol Q\cdot\mathbf n_r\Big]dS.
\end{aligned}
\label{eq:s-adjoint-remainder}
\end{equation}

The outward normal on the inner boundary is \(-\mathbf n_r\), so the
contact terms involving \(\boldsymbol Q\) cancel. If the remote
remainder vanishes, substitution of Eq.~\eqref{eq:s-contact-flux} gives

\begin{equation}
\begin{aligned}
J_{xy}&=2\int_{\mathcal P}\boldsymbol Q\cdot\nabla_rw\,d^3r\\
&=-8\int_{\mathcal P}d^3r\oint_{|2\mathbf n-\mathbf r|>2}
g_3(\mathbf r,2\mathbf n)\,\mathbf n\cdot\nabla_rw\,d\Omega_n.
\end{aligned}
\label{eq:s-adjoint-stress}
\end{equation}

At zero shear write \(w=w_0+\mathrm{Pe}\,w_1+\cdots\), with
\(w_0=\psi_2/8\). The first correction solves
\(2\nabla_r^2w_1=-r_y\partial_xw_0\) with homogeneous contact
derivative. Its source has zero angular monopole:

\begin{equation}
\overline{r_y\partial_x(xy/r^5)}
=r^{-3}\left(\overline{n_{r,y}^2}-5\overline{n_{r,x}^2n_{r,y}^2}\right)
=r^{-3}\left(\frac13-\frac5{15}\right)=0.
\label{eq:s-zero-monopole}
\end{equation}

The decaying homogeneous monopole is zero. Since the equilibrium flux
\(\boldsymbol Q_0\) is radial, its integral against \(\nabla_rw_1\)
vanishes. The linear part of Eq.~\eqref{eq:s-adjoint-stress} therefore
uses only \(\psi_3\) and \(w_0\). The required condition is the
vanishing of the first-order remote remainder, together with the density
and shear limits used in this reduction.

\subsection{S1.3 Exact superposition
cancellation}\label{s1.3-exact-superposition-cancellation}

Let \(\boldsymbol\Psi_k\) be the exclusion flux built from \(H\) for the
\(k\)th pair of the triplet, and put

\begin{equation}
T_k=\int_{r>2}\boldsymbol\Psi_k(\mathbf r)\cdot\nabla H(\mathbf r)\,d^3r,
\qquad S=T_1+T_2+T_3.
\label{eq:s-three-pair-integrals}
\end{equation}

With \(\psi_2=-(16/3)H\) and \(w_0=-(2/3)H\), \(J_1^K=(64/9)S\) and
\(c_3^K=-(72/\pi^2)S\). The first-pair flux is \(H\) times the radial
equilibrium exclusion flux, of magnitude \(4\pi(1-r^2/16)\) on
\(2<r<4\). The second-pair term integrates the contact angular function
over the accessible cap. For the third pair the harmonic mean-value
property gives a full-sphere contribution \(4\pi^2/15\) and the excluded
cap subtracts \(9\pi^2/40\). The results are

\begin{equation}
T_1=-\frac{9\pi^2}{320},\qquad
T_2=-\frac{13\pi^2}{960},\qquad
T_3=\frac{\pi^2}{24},\qquad S=0.
\label{eq:s-cancellation-values}
\end{equation}

The full-sphere and excluded-cap pieces of \(T_3\) use the contact-flux
integral at fixed \(r\) before the radial integration. Individually they
are conditionally convergent as five-dimensional integrals; this
integration order is part of their definition.

An elementary radial check makes the cancellation explicit. Put
\(W=1-r^2/16\). After the angular integral,

\begin{equation}
\begin{aligned}
T_1&=-\frac{16\pi^2}{5}\int_2^4 W r^{-5}\,dr,\\
T_2+T_3&=\frac{\pi^2}{10}\int_2^4 W(r^{-2}+3/16)\,dr.
\end{aligned}
\label{eq:s-radial-cancellation}
\end{equation}

The primitives \(F_1=r^{-4}(1-r^2/16)^2\) and \(F_2=r^{-1}(1-r^2/16)^2\)
satisfy \(F_1'=-4Wr^{-5}\) and \(F_2'=-W(r^{-2}+3/16)\). Both vanish at
four and \(F_1(2)=F_2(2)/8\), proving the zero without fitting a
coefficient.

The contact theorem gives a second interpretation. Let \(B\) be the
excluded-volume ball of radius two and \(V(r)=(\pi/12)(8+r)(4-r)^2\) its
overlap volume with a translated copy, for \(r\le4\). If \(\Phi_B\) is
its Newtonian potential and
\(E=(\mathbf e_x\mathbf e_y+\mathbf e_y\mathbf e_x)/2\), then
\(H=E:\nabla\nabla\Phi_B/(3|B|)\) outside \(B\), while the traceless
Hessian is zero inside. For a touching center \(|\mathbf x|=2\), put
\(X_{\mathbf x}=\{\mathbf y:|\mathbf y|>2,|\mathbf y-\mathbf x|<2\}\).
Newton's shell theorem gives

\begin{equation}
\begin{aligned}
\int_{X_{\mathbf x}}H(\mathbf y)\,d^3y&=\kappa_2 H(\mathbf x),\\
\kappa_2&=\frac{M_{\rm lens}}{|B|}-V(2)
=\frac{3B_3}{2B_2}-V(2).
\end{aligned}
\label{eq:s-lens-multiplier}
\end{equation}

Here \(M_{\rm lens}=\int_BV\,d^3y=3B_3\) and \(|B|=2B_2\). The
equilibrium contact theorem gives \(B_3=B_2V(2)=5B_2^2/8\), with
\(V(2)=10\pi/3\), and hence \(\kappa_2=V(2)/2\). Gauss's law at fixed
separation of the second and third spheres then gives

\begin{equation}
\begin{aligned}
T_3&=\frac32\oint_{|\mathbf x|=2}H(\mathbf x)
\int_{X_{\mathbf x}}H(\mathbf y)\,d^3y\,dS_x\\
&=\kappa_2\int_{r>2}|\nabla H|^2\,d^3r=\frac{\pi^2}{24}.
\end{aligned}
\label{eq:s-third-pair-contact}
\end{equation}

Equivalently, integrating the superposition sum over the third sphere in
a linear-response cluster expansion of the form used in {[}4{]} gives the
contact density correction

\begin{equation}
\begin{aligned}
g^{(1)}_{\rm cl}(\mathbf x)
&=\mathrm{Pe}\left[V(2)\psi_2(\mathbf x)
-2\int_{X_{\mathbf x}}\psi_2(\mathbf y)\,d^3y\right]\\
&=\mathrm{Pe}[V(2)-2\kappa_2]\psi_2(\mathbf x)=0.
\end{aligned}
\label{eq:s-cluster-contact}
\end{equation}

The equilibrium contact-density rise contributes \(+6\) to \(c_3\) and
the two other-pair perturbations contribute \(-3\) each. This is the
contact cancellation identified in Lionberger's low-density
hypernetted-chain calculation {[}5, Eq. (4.107), Eqs. (A.60)--(A.62){]}.
It follows from the geometry and contact theorem; it does not assert
that superposition satisfies the three-body contact condition.

\subsection{S1.4 Closure and stress
comparisons}\label{s1.4-closure-and-stress-comparisons}

At this order the hypernetted-chain closure {[}6, Eq. (35){]}, {[}7,
Eqs. (31), (32), (79), (90){]}, Szamel's nondecoupled closure {[}8, Eqs.
(11), (12){]} and equilibrium-corrected superposition {[}9, Eqs. (2.20),
(2.25), (2.26){]} retain the same three pair terms with unit weights.
Ohtsuki's superposition reduction does so as well {[}10, Eqs.
(2.8)--(2.11){]}. Their common contact-stress coefficient is therefore
zero. The closure weights and their common-stress coefficients are
compared in Table 3 of the main article. These reductions concern the
no-hydrodynamic-interaction dilute limit, rather than
finite-concentration results in the cited papers.

\textbf{Joint versus conditional equilibrium.} A joint-equilibrium
three-body prescription {[}11, Eqs. (26), (27){]} has no first-order
shear response in \(g_3\), but retains the equilibrium pair correction
\(V(r)\). Substitution of
\(g^{(1)}=V+\mathrm{Pe}\,A(r)n_{r,x}n_{r,y}+O(\mathrm{Pe}^2)\) into
Eq.~\eqref{eq:s-pair-hierarchy} gives

\begin{equation}
\begin{aligned}
A''+\frac2rA'-\frac6{r^2}A&=\frac r2 V', &&r>2,\\
A'(2)&=V(2),\qquad A(\infty)=0.
\end{aligned}
\label{eq:s-equilibrium-closure-ode}
\end{equation}

Outside the source shell the decaying solution is proportional to
\(r^{-3}\). The equation implies \(\partial_r(A'/r+3A/r^2)=V'/2\),
whence

\begin{equation}
\begin{aligned}
\left[\frac{A'}r+\frac{3A}{r^2}\right]_2^\infty
&=\frac12[V]_2^\infty,\\
\frac{A'(2)}2+\frac{3A(2)}4&=\frac{V(2)}2,\qquad A(2)=0.
\end{aligned}
\label{eq:s-equilibrium-closure-contact}
\end{equation}

Thus the contact shear projection is \((4\pi/15)A(2)=0\). This is a
three-dimensional reduction of the joint prescription; it is distinct
from the conditional-equilibrium prescription in Eq. (29) of Tschopp and
Brader {[}11{]}. The latter retains the first-pair shear response in the
joint distribution and gives the Russel--Gast reduction {[}12{]}.

\textbf{Contact density and stress prescription.} Write
\(g=g_{\rm eq}[1-\tfrac12\mathrm{Pe}\,n_xn_y(f_0+\phi f_1+\cdots)]+O(\mathrm{Pe}^2)\),
where \(f_0=(32/3)r^{-3}\). Outside contact
\(g_{\rm eq}=1+\phi g_1+\cdots\), with \(g_1=8-3r+r^3/16\) for \(2<r<4\)
and zero beyond. The contact value \(g_1(2)=5/2\) gives

\begin{equation}
c_3=\frac95\left[g_1(2)f_0(2)+f_1(2)\right]
=6+\frac95f_1(2).
\label{eq:s-contact-density-reduction}
\end{equation}

For Russel--Gast, the structure solution \(f_1(2)=-53/24\) gives
\(81/40\). Using its mean-force stress instead removes the
contact-density contribution and gives \(-159/40\) {[}12, Eqs. (29),
(30), (42){]}, {[}1, Eqs. (3.15), (3.20), (3.21){]}. This comparison
uses that known structure solution as input; it compares stress
prescriptions. For superposition \(f_1(2)=-10/3\), so \(6-6=0\),
consistent with Eq.~\eqref{eq:s-cancellation-values}. Szamel's
additional decoupling retains the explicit contact-density contribution
and the third-pair term {[}8, Eqs. (19), (23)--(25){]}, giving
\(6-(72/\pi^2)T_3=3\). Keeping only the contact-density rise gives six,
the Pair prescription {[}7, Eq. (29){]} obtained from contact-value
scaling {[}13, Eqs. (53), (54){]} when hydrodynamic interactions are
removed. These contact-density groupings are not additional sets of
three pair weights.

For the bare-potential Percus--Yevick closure, replacing the first pair
by its Boltzmann factor removes \(T_1\) {[}14, Eqs. (16), (46){]}, {[}6,
Eq. (36){]}. Its contact reduction is \(-81/40\). The thesis values
\(-8.025\) for this closure and 5.0 for the Pair prescription {[}5, Eqs.
(4.109), (4.110){]} differ from the respective common-stress reductions
\(-2.025\) and six; the latter follow from the equations and contact
factor above. These reported values are distinct from the contact-stress
reductions derived here.

\section{S2. Weak correction and an explicit two-sided
certificate}\label{s2.-weak-correction-and-an-explicit-two-sided-certificate}

\subsection{S2.1 Energy space and trace
estimate}\label{s2.1-energy-space-and-trace-estimate}

The center-of-mass representative of the relative coordinates is

\begin{equation}
\begin{aligned}
\mathbf x_1&=-(\mathbf r+\mathbf s)/3,\\
\mathbf x_2&=(2\mathbf r-\mathbf s)/3,\\
\mathbf x_3&=(2\mathbf s-\mathbf r)/3,\\
\varrho^2&=\sum_i|\mathbf x_i|^2=\mathcal R^2/3,\qquad
\mathcal R^2=r^2+s^2+q^2.
\end{aligned}
\label{eq:s-cm-coordinates}
\end{equation}

The Euclidean metric of \(X=(\mathbf x_1,\mathbf x_2,\mathbf x_3)\) on
\(\sum_i\mathbf x_i=0\) has inverse gradient metric
\(\left(\begin{smallmatrix}2&1\\1&2\end{smallmatrix}\right)\otimes I_3\)
in \((\mathbf r,\mathbf s)\). Since \(\nabla_1=-\nabla_r-\nabla_s\),
\(\nabla_2=\nabla_r\) and \(\nabla_3=\nabla_s\), define

\begin{equation}
\mathcal D[f,g]=\iint_\Omega\sum_{i=1}^3\nabla_i f\cdot\nabla_i g\,d^3r\,d^3s.
\label{eq:s-dirichlet-form}
\end{equation}

The volume element is \(dV_X=c_0d^3r\,d^3s\), \(c_0=3^{-3/2}\). On the
accessible contact face \(\Sigma_{12}\), the inward unit normal is
\(\nu=(-\mathbf n,\mathbf n,0)/\sqrt2\). The full separating derivative
is \(\partial=(\nabla_2-\nabla_1)\cdot\mathbf n\), so
\(\partial_\nu=\partial/\sqrt2\), and

\begin{equation}
dS_X=\sqrt2c_0dS_r\,d^3s,\qquad
X\cdot\nu=\sqrt2,\qquad dS_r=4d\Omega_r.
\label{eq:s-contact-measures}
\end{equation}

The other two faces follow by relabeling. Let \(\mathcal C\) consist of
restrictions to \(\Omega\) of ambient smooth compactly supported
functions invariant under all six label permutations. They need not
vanish on contact. The energy space \(\mathcal H\) is their completion
in \(\mathcal D^{1/2}\). Contact intersections cause no extra boundary
integral: the dilation field \(X\) is transverse to each active contact
surface, so a truncated domain is locally above the maximum of smooth
graphs and is Lipschitz; the intersections have zero surface measure.

For \(f\in\mathcal C\) put \(\mathcal K=\int|\nabla_Xf|^2dV_X\),
\(\mathcal V=\int f^2/\varrho^2dV_X\) and
\(\mathcal S=\sum_{i<j}\int_{\Sigma_{ij}}\sqrt2f^2/\varrho^2dS_X\). The
divergence identity \(\nabla_X\cdot(X/\varrho^2)=4/\varrho^2\) gives,
for every real \(t\),

\begin{equation}
0\le\int_\Omega\left|\nabla_Xf+t f\frac{X}{\varrho^2}\right|^2dV_X
=\mathcal K-t\mathcal S-t(4-t)\mathcal V.
\label{eq:s-hardy-square}
\end{equation}

In particular \(\mathcal S\le\mathcal K/4\) at \(t=4\) and
\(4\mathcal V\le\mathcal K\) at \(t=2\). By permutation symmetry and
Eq.~\eqref{eq:s-contact-measures}, \(\mathcal S=18c_0I(f)\), where
\(I(f)=\int_{\Sigma_{12}}f^2/\mathcal R^2\,dS_r\,d^3s\), while
\(\mathcal V=3c_0\int_\Omega f^2/\mathcal R^2\) and
\(\mathcal K=c_0\mathcal D[f,f]\). Hence

\begin{equation}
72I(f)\le\mathcal D[f,f],\qquad
\iint_\Omega\frac{f^2}{\mathcal R^2}\,d^3r\,d^3s
\le\frac1{12}\mathcal D[f,f].
\label{eq:s-trace-inequality}
\end{equation}

The interior bound makes the energy a norm and realizes its completion
as functions with square-integrable gradients and weighted
square-integrable values. The contact inequality extends by continuity
and defines their weighted mean-square traces. It makes no pointwise
assertion at a contact intersection. The constant 72 is sharp on
\(\mathcal H\): \(f=\mathcal R^{-4}\) makes the square at \(t=4\)
vanish, and ambient smooth outer cutoffs approximate it with energy
error \(O(L^{-4})\). Its possible improvement on the shear subspace is
not needed below.

\subsection{S2.2 Weak correction, positivity and physical
conditions}\label{s2.2-weak-correction-positivity-and-physical-conditions}

The superposition contact error on \(\Sigma_{12}\) is

\begin{equation}
e=\mathbf n\cdot[\nabla\psi_2(\mathbf s)-\nabla\psi_2(\mathbf q)],
\qquad \mathbf r=2\mathbf n.
\label{eq:s-contact-error}
\end{equation}

Define the single-contact functional
\(L_e(f)=\int_{\Sigma_{12}}fe\,dS_r\,d^3s\). Cauchy--Schwarz and
Eq.~\eqref{eq:s-trace-inequality} give

\begin{equation}
|L_e(f)|\le M\mathcal D[f,f]^{1/2},\qquad
M^2=\frac1{72}\int_{\Sigma_{12}}e^2\mathcal R^2\,dS_r\,d^3s<\infty.
\label{eq:s-bounded-source}
\end{equation}

The source is bounded on bounded contact sets. For a remote third sphere
its two gradients cancel at leading order, giving \(e=O(s^{-5})\) and
\(e^2\mathcal R^2=O(s^{-8})\), integrable over \(d^3s\). Each gradient
separately also defines a bounded pairing, since it is \(O(s^{-4})\).

The correction \(\chi=\psi_3-\psi_3^K\) has \(\mathscr L\chi=0\) in
\(\Omega\), with \(\mathscr L=\sum_i\Delta_i\), and full inward
separating derivative \(\partial\chi=-e\) on each contact. Green's
identity and Eq.~\eqref{eq:s-contact-measures} therefore give the weak
problem

\begin{equation}
\mathcal D[f,\chi_w]=3L_e(f)\quad(f\in\mathcal H).
\label{eq:s-weak-correction}
\end{equation}

The three counts unordered faces; \(L_e\) is a single-face integral.
The source is bounded, so the Riesz representation theorem supplies exactly one \(\chi_w\in\mathcal H\). Taking
\(f=\chi_w\) and completing the square give

\begin{equation}
\begin{aligned}
\mathcal E[\chi_w]&=\frac16\mathcal D[\chi_w,\chi_w]
=\frac12L_e(\chi_w),\\
\mathcal E[\chi_w]-\left[L_e(f)-\frac16\mathcal D[f,f]\right]
&=\frac16\mathcal D[f-\chi_w,f-\chi_w].
\end{aligned}
\label{eq:s-weak-energy}
\end{equation}

This energy is strictly positive. Indeed, \(e\) is continuous and
nonzero at some contact point away from the other faces. An ambient
smooth test supported near that point, chosen to have the sign of \(e\)
and summed over label permutations, has positive \(L_e\). Thus the
source functional and its Riesz representative are nonzero. This proves
positivity without a closure approximation.

For completeness, the stress normalization follows from relabeling the
two other-pair contributions to \(e\). If \(e_0\) denotes one, then
\(L_e(\chi)=2\langle\chi,e_0\rangle\) and the adjoint stress gives
\(J_{1,\chi}=-\langle\chi,e_0\rangle/4\). Since superposition
contributes zero,

\begin{equation}
c_{3,\chi}=-\frac{81}{8\pi^2}J_{1,\chi}
=\frac{81}{32\pi^2}\mathcal E[\chi].
\label{eq:s-energy-stress}
\end{equation}

Set \(K=81/(32\pi^2)\) and \(\Theta=K\mathcal E[\chi_w]\). All weak
inequalities below concern \(\Theta\). Their physical interpretation
\(c_3(0)=\Theta\) requires both conditions of Proposition 1 in the main
article: the zero-shear physical three-body correction exists and
selects \(\chi_w\), including the remote cluster behavior implicit in
that selection; and the first-order pair-density adjoint remainder in
Eq.~\eqref{eq:s-adjoint-remainder} vanishes in the density and shear
limits. A sufficient criterion for the first condition is that the
physical first-order correction belongs to the completion
\(\mathcal H\), including all remote cluster ends, and satisfies
Eq.~\eqref{eq:s-weak-correction} for every compact ambient test.
Continuity of the source and density of compact tests extend this
identity to \(\mathcal H\); uniqueness then identifies the physical
correction with \(\chi_w\).

\subsection{S2.3 Global trials and exact
integration}\label{s2.3-global-trials-and-exact-integration}

For a pair separation \(r_{ij}\) and its unit direction
\(\mathbf n_{ij}\), use the six-term symmetric modes

\begin{equation}
\begin{aligned}
V_{\lambda\mu\nu\tau}
&=\sum_{\substack{i,j,k\\\mathrm{distinct}}}
(2/r_{ij})^{\lambda+1+\mu}(2/r_{ik})^\nu(2/r_{jk})^\tau\\
&\hspace{12mm}\times(\mathbf n_{ij}\cdot\nabla_{\mathbf z})^\lambda
\psi_2(\mathbf z)\big|_{\mathbf z=\mathbf x_k-\mathbf x_i}.
\end{aligned}
\label{eq:s-global-mode}
\end{equation}

The direction is held fixed in this derivative; only \(\psi_2\) is
differentiated. Every mode is smooth near the closed accessible domain.
Its homogeneous degree is \(k=2\lambda+4+\mu+\nu+\tau\). The 500-mode
lower field uses the catalogue in Section S2.5. Its modes have finite
energy on every remote cluster end, and smooth outer cutoffs converge in
energy, placing the field in \(\mathcal H\).

The 496-mode upper field sets four lower-only coefficients,
\((\lambda,\mu,\nu,\tau)=(0,1,j,0)\) for \(j=0,1,2,3\), to zero. These
four have finite energy but their weighted residuals are unsuitable for
the upper comparison. Remote admissibility cannot be checked only by
simultaneous scaling of all sides. Here \(R\) denotes the distance of a
remote spectator from a pair whose separation stays bounded. For
\(\lambda=1,\mu=\tau=0\), the two orderings whose derivative acts on
that close pair have prefactors of order \(R^{-2}\). Their leading terms
cancel: the gradient of the even pair response is odd under
reversal of the close-pair separation. The combined contribution starts
at \(R^{-3}\). The orderings whose derivative acts on a remote pair
already decay at least as \(R^{-4}\). The \(\lambda=0,\mu=1,\tau=0\)
modes lack this odd-parity cancellation. For the listed upper field, the
contact and bulk residuals have the weighted square integrability needed
in Section S2.4. This check is a sufficient verification for these
specified modes, not a necessary-and-sufficient criterion for all
possible linear combinations.

Let \(B_i=L_e(V_i)\) and \(D_{ij}=\mathcal D[V_i,V_j]\). An exact
rational vector \(c\) defines \(f=\sum_i c_iV_i\) and

\begin{equation}
F(f)=K\left(c^TB-\frac16c^TDc\right)\le\Theta.
\label{eq:s-trial-functional}
\end{equation}

This inequality evaluates the actual field, independently of whether it
maximizes the functional in the chosen span.

Rotational reduction uses \(E_{xy}=E_{yx}=1/2\), \(E:E=1/2\), and, for
symmetric tensors \(A,B\),

\begin{equation}
\overline{(E:A)(E:B)}=\frac{E:E}{5}
\left(A:B-\frac{\operatorname{tr}A\operatorname{tr}B}{3}\right).
\label{eq:s-rotation-average}
\end{equation}

The side-length volume measure is \(8\pi^2rsq\,dr\,ds\,dq\). Order the
sides as \(\ell(1,u,v)\), with \(\ell\ge2\) and \(1\le u\le v\le u+1\);
all six orderings cover the domain. On a contact face put \(s=2u,q=2v\).
The measure is \(256\pi^2uv\,du\,dv\), retaining both orderings of the
other sides. Form Cartesian gradients before averaging orientations, so
all shape-gradient terms are retained. For mode degrees \(k,k'\) the
scale integral gives

\begin{equation}
\begin{aligned}
D_{ij}&=\frac{48\pi^2}{10}\frac{2^{4-k-k'}}{k+k'-4}
\int_1^\infty du\int_u^{u+1}dv\,uv\,G_{ij}(u,v),\\
B_i&=\frac{2\cdot256\pi^2}{10}2^{-k-4}
\int_1^\infty du\int_u^{u+1}dv\,uv\,P_i(u,v),
\end{aligned}
\label{eq:s-global-integrals}
\end{equation}

The reduced contractions are defined directly from the modes by

\begin{equation}
\begin{aligned}
G_{ij}(u,v)&=10\left[\ell^{k+k'+2}
\overline{\sum_{m=1}^3\nabla_mV_i\cdot\nabla_mV_j}\right]_{\ell=2},\\
P_i(u,v)&=10\,2^{k+4}\left[\overline{V_i e}\right]_{r=2,\,s=2u,\,q=2v}.
\end{aligned}
\label{eq:s-reduced-contractions}
\end{equation}

The bar denotes the orientation average in
Eq.~\eqref{eq:s-rotation-average}, with the triangle shape fixed.
Homogeneity makes the first bracket independent of the positive scale
\(\ell\). These definitions include the shear normalization; the factors
outside the integrals in Eq.~\eqref{eq:s-global-integrals} account for
side permutations and contact radius.

The integrands are Laurent polynomials. Integration in \(v\) is
elementary; \(v^{-1}\) generates \(\ln(1+1/u)\). Partial fractions and
integration by parts handle the remaining integral, including
\(\int_1^\infty u^{-1}\ln(1+1/u)du=\pi^2/12\). Divergent terms are
combined before the remote limit. Each resulting entry has the form
\(\pi^2(x_0+x_1\ln2)+\pi^4x_2\), with rational \(x_j\). Exact
coefficients and outward rational bounds on \(\pi\) and \(\ln2\) then
enclose Eq.~\eqref{eq:s-trial-functional}. The ray expression also uses
\(\sqrt{10}\), bounded by rational endpoints whose squares bracket ten.

\subsection{S2.4 Radial residual enclosure in the shear
sector}\label{s2.4-radial-residual-enclosure-in-the-shear-sector}

For the functional \(F\) in Eq.~\eqref{eq:s-trial-functional}, define
the residual

\begin{equation}R_v(f)=3L_e(f)-\mathcal D[v,f].
\label{eq:s-ray-min-residual}\end{equation}

Equation~\eqref{eq:s-weak-energy} gives
\(\Theta-F(v)=K\|\chi_w-v\|_{\mathcal D}^2/6=K\|R_v\|_{\mathcal D^*}^2/6\).
An upper bound on this residual norm therefore supplies an upper
enclosure.

Rotations commute with the energy operator, permutation symmetry and the
contact source. The source is linear in a traceless shear tensor \(E\),
so the solution and the chosen upper field belong to angular degree
\(l=2\). The \(l=2\) orthogonal projection commutes with the energy form
and leaves the residual functional unchanged, so its full dual norm may
be evaluated in this sector. Finiteness of the weighted volume and
contact norms defined in the main article makes both source pairings
continuous on \(\mathcal H\), by Eq.~\eqref{eq:s-trace-inequality}. This
extends the residual identity from compact tests to energy-space tests.
For the physical rotation generators
\(\mathcal J_a=\sum_i(\mathbf e_a\times\mathbf x_i)\cdot\nabla_i\), the
Casimir identity on each rotation orbit is
\(\sum_a\|\mathcal J_a f\|^2=6\|f\|^2\). Moreover
\(\sum_a|\mathcal J_a f|^2\le\varrho^2|\nabla_{\rm tan}f|^2\), where
\(\varrho^2=\sum_i|\mathbf x_i|^2=(r^2+s^2+q^2)/3\). Indeed, the Gram
matrix of the infinitesimal rotation directions is
\(\varrho^2I-\sum_i\mathbf x_i\mathbf x_i^T\preceq\varrho^2I\), which
gives this gradient inequality. After orientation integration, the
radial and tangential energies therefore bound the energy below by a
radial derivative term plus \(6f^2/\varrho^2\). The remaining
nonnegative shape-gradient energy can be discarded.

Order the three side lengths as \(\ell,\ell u,\ell v\), with
\(\ell\ge2\), \(1\le u\le v\le u+1\), and put \(R_*^2=1+u^2+v^2\). Bold
amplitudes below are symmetric traceless body tensors: the scalar field
is \(E:Q\widehat f Q^T\), with \(E:E=1/2\), and its Haar-averaged
product is the Frobenius product of amplitudes divided by ten. The six
side orderings give the reduced measure
\(d\zeta=(48\pi^2/10)uv\,du\,dv\). At each shape the retained energy is

\begin{equation}
\begin{aligned}
q_6[f]&=\frac3{R_*^2}\int_2^\infty
 \left(\ell^5|\partial_\ell f|^2+6\ell^3|f|^2\right)d\ell\\
&=\frac3{R_*^2}
 \left\{\int_0^\infty\!\left(|w'|^2+10|w|^2\right)dx
       +2|w(0)|^2\right\},\\
\ell&=2e^x,\qquad f=\ell^{-2}w.
\end{aligned}
\label{eq:s-ray-min-energy}
\end{equation}

The endpoint term follows by integrating \(-4w'w\); it has the positive
sign shown. The identity first holds for compact radial tests and
extends to their half-line energy completion. Enlarging the original
test space to these independent radial functions can only increase the
residual dual norm.

Let \(\mathscr L=\sum_i\Delta_i\), and use the full separating
derivative \(\partial v=(\nabla_2-\nabla_1)v\cdot\mathbf n\) on
\(\Sigma_{12}\). Green's identity gives
\(R_v(f)=\int_\Omega f\mathscr Lv+3\int_{\Sigma_{12}}fb_v\), where
\(b_v=e+\partial v\). Write the homogeneous bulk residual as
\(\mathscr Lv=\sum_k\ell^{-k-2}\boldsymbol G_k\), and set \(p_k=k-2\),
\(\boldsymbol a_k=2^{2-k}\boldsymbol G_k
=16(\mathscr Lv)_k|_{\ell=2}\). After the orientation average and the
two orderings of the other sides, the single-face contact measure is
\((512\pi^2/10)uv\,du\,dv=(32/3)d\zeta\). The three faces and
\(f(2)=w(0)/4\) give the boundary-source factor \(3(32/3)/4=8\). The ray
functional is therefore

\begin{equation}
R_v(f)=\int d\zeta\left\{
 \int_0^\infty\sum_k\boldsymbol a_k e^{-p_kx}:w(x)\,dx
 +8\boldsymbol b_v:w(0)\right\}.
\label{eq:s-ray-min-source}
\end{equation}

Put \(\omega=\sqrt{10}\). The inverse of \(-d^2/dx^2+\omega^2\) with
Robin condition \(w'(0)=2w(0)\) has endpoint kernel
\(G_\omega(0,0)=1/(\omega+2)\). For a bulk source \(e^{-qx}\), its
response is \([e^{-qx}-(q+2)e^{-\omega x}/(\omega+2)]/(\omega^2-q^2)\).
Integrating against \(e^{-px}\) gives

\begin{equation}
I_\omega(p,q)=
 \frac{p+q+\omega+2}
 {(p+q)(\omega+p)(\omega+q)(\omega+2)},\qquad
\int_0^\infty e^{-px}G_\omega(x,0)\,dx
 =\frac1{(\omega+2)(\omega+p)}.
\label{eq:s-ray-min-kernel}
\end{equation}

Consequently the enlarged squared residual norm and the enclosure are

\begin{equation}
\begin{aligned}
\eta(v)
&=\int d\zeta\,\frac{R_*^2}{3}
 \left[
 \sum_{k,l}(\boldsymbol a_k:\boldsymbol a_l)I_\omega(p_k,p_l)
 +16\sum_k\frac{\boldsymbol a_k:\boldsymbol b_v}
                 {(\omega+2)(\omega+p_k)}
 +\frac{64|\boldsymbol b_v|^2}{\omega+2}
 \right],\\
F(f_{\rm L})&\le\Theta\le F(v)+\frac K6\eta(v).
\end{aligned}
\label{eq:s-ray-min-enclosure}
\end{equation}

The prefactor \(R_*^2/3\) is the inverse of the energy weight in
Eq.~\eqref{eq:s-ray-min-energy}, and the contact term contains the
square \(8^2=64\) of the boundary-source factor. The mixed term is
retained with its sign; although it may be negative, the complete
bracket is a positive Green quadratic form. The coefficients specified
in Section S2.5 determine both fields exactly. Their reduced integrands
are Laurent polynomials; the integration rules of Section S2.3 give
rational combinations of \(1,\ln2,\pi^2\), and the upper expression also
contains \(\sqrt{10}\). Outward rational bounds on these constants
give the evaluated enclosures in Table S2.

\subsection{S2.5 Complete trial specification and evaluated
enclosure}\label{s2.5-complete-trial-specification-and-evaluated-enclosure}

Table S1 specifies the lower field \(f_{\rm L}=\sum_{i=1}^{500}c_iV_i\)
and the upper field \(v=\sum_{i=1}^{500}d_iV_i\) in a common ordered
catalogue. The Lower and Upper columns give the numerators of \(c_i\)
and \(d_i\), respectively. The four modes omitted from the upper
catalogue have \(d_i=0\), leaving 496 modes in that catalogue. The
four-digit code \(1000\lambda+100\mu+10\nu+\tau\) identifies each
tuple, with leading zeros retained.

\begingroup
\fontsize{10}{12}\selectfont
\setlength{\tabcolsep}{3pt}
\setlength{\LTpre}{6pt}
\setlength{\LTpost}{8pt}
\begin{longtable}{@{}rrr@{\hspace{16pt}}rrr@{}}
\caption{Exact global trial fields. Coefficients are the listed numerators divided by $2^{32}$.}\label{tab:s-trial-fields}\\
\toprule
Mode & Lower & Upper & Mode & Lower & Upper\\
\midrule
\endfirsthead
\multicolumn{6}{c}{Table S1. Exact global trial fields (continued).}\\
\toprule
Mode & Lower & Upper & Mode & Lower & Upper\\
\midrule
\endhead
\midrule
\multicolumn{6}{r}{Continued on next page}\\
\endfoot
\bottomrule
\endlastfoot
\texttt{0100} & -7908319 & 0 & \texttt{1124} & 1401661915117574 & -151463061861751\\
\texttt{0101} & 798585525 & 5324907 & \texttt{1204} & -291844387813251 & -95790530146987\\
\texttt{0102} & -27214479676 & 963914555 & \texttt{1214} & 952169179497681 & 1308801624793707\\
\texttt{0110} & 31752755 & 0 & \texttt{1224} & -728858928914583 & -620871824599256\\
\texttt{0111} & -13601246747 & -2420908742 & \texttt{2023} & -82162209628966 & -145135868211651\\
\texttt{0112} & 613762809520 & 37490140282 & \texttt{2123} & -316902648841466 & -163629677895024\\
\texttt{0120} & -42043822 & 0 & \texttt{2203} & -223625197788145 & -154269921046447\\
\texttt{0121} & 31881815885 & 33619630978 & \texttt{2213} & -414849378393115 & -213951646689832\\
\texttt{0122} & -7838255664767 & -2061839745396 & \texttt{2223} & -484966052746254 & -2855361806177\\
\texttt{0130} & 18045943 & 0 & \texttt{2024} & 204426535305920 & 186518406160929\\
\texttt{0131} & 986133092816 & 199463489295 & \texttt{2124} & 679114454036447 & -124121179472357\\
\texttt{0132} & 11894571695733 & 3465572657117 & \texttt{2204} & 661038309891006 & 525683044911489\\
\texttt{0200} & 93923018 & 3622087 & \texttt{2214} & 775441539555726 & -111566059643477\\
\texttt{0201} & -18363574084 & -3218155309 & \texttt{2224} & 408357884993487 & 18278594970775\\
\texttt{0202} & 607220767039 & 195299997833 & \texttt{3223} & -54250675922979 & -30964932598392\\
\texttt{0210} & 10064459114 & 2392652078 & \texttt{3024} & -58974773861130 & -44036694582011\\
\texttt{0211} & -233504561540 & -109735917928 & \texttt{3124} & -10806110394555 & -50376003460330\\
\texttt{0212} & -9273144682891 & -1340740988954 & \texttt{3204} & -271115655734208 & -203469494423241\\
\texttt{0220} & -27105065602 & -33583052532 & \texttt{3214} & 148554953443138 & 109601152637667\\
\texttt{0221} & 7941839603269 & 2849149624779 & \texttt{3224} & 30163666602468 & 11670057679096\\
\texttt{0222} & 59192678095584 & 6433823281009 & \texttt{4220} & -1007003255471 & -997729780613\\
\texttt{0230} & -988227850272 & -199479932532 & \texttt{4221} & 19621166605719 & 7615159864482\\
\texttt{0231} & -24538744578736 & -5948733347555 & \texttt{4222} & -50639523436493 & -11856067098559\\
\texttt{0232} & -4053279356278 & -10480510545153 & \texttt{4023} & 14296390110739 & 2689973123014\\
\texttt{0300} & -68444392 & -3347660 & \texttt{4123} & -29653711788812 & 3999666108692\\
\texttt{0301} & 39238958998 & -52837439654 & \texttt{4203} & -23477242787755 & -26813858031160\\
\texttt{0302} & -276595346249 & 783301506454 & \texttt{4213} & 35269850922561 & 51855349720579\\
\texttt{0310} & -33315987124 & 53577397654 & \texttt{4223} & 28043217114287 & -16371498825460\\
\texttt{0311} & 792040437724 & -444604612681 & \texttt{3005} & -242254409522663 & -441545960571333\\
\texttt{0312} & 5648892864746 & -2445505619713 & \texttt{3015} & -642674446099955 & -23072965577789\\
\texttt{0320} & -611214739871 & -353562550809 & \texttt{3105} & -858983667042410 & -388617144165670\\
\texttt{0321} & -22063807158454 & -391270630744 & \texttt{3115} & 696810415589724 & 313202364566665\\
\texttt{0322} & -20880933376226 & -44447189512258 & \texttt{4004} & 35551609708418 & 45938388034019\\
\texttt{0330} & 10668838566764 & 1408027660908 & \texttt{4014} & 91807498232307 & 10087551425551\\
\texttt{0331} & -5272436960601 & 12829708940750 & \texttt{4024} & -55851411664663 & -16423596232693\\
\texttt{0332} & 5927577518835 & 14261620273904 & \texttt{4104} & 39257747678279 & -26570591758546\\
\texttt{1000} & -303275177 & -35349498 & \texttt{4114} & -67336224542606 & 46713764607756\\
\texttt{1001} & -53962202957 & -3339753963 & \texttt{4124} & 45302042993648 & -8892515431172\\
\texttt{1002} & 763705589266 & 294321125407 & \texttt{4204} & -11795323788777 & 21035752038891\\
\texttt{1010} & 1010155246 & 147959964 & \texttt{4214} & -24501389269664 & -70388200544148\\
\texttt{1011} & 677032814669 & -57404462552 & \texttt{4224} & 13578186494000 & 42917207598282\\
\texttt{1012} & -14283457314267 & -3566714770543 & \texttt{3025} & 226227600853040 & 53583561004205\\
\texttt{1020} & -1128260501 & -185938689 & \texttt{3125} & -289329659648084 & -42492680224225\\
\texttt{1021} & -1016570184197 & 1166811366787 & \texttt{3205} & 473825774137031 & 371219552894745\\
\texttt{1022} & 57921508441030 & 6954054725367 & \texttt{3215} & -542940124657309 & -388957056245674\\
\texttt{1030} & 418257522 & 75402046 & \texttt{3225} & 160921525124064 & 98870943283547\\
\texttt{1031} & -5418286177281 & -2871921674030 & \texttt{1025} & -812902079598452 & -91554788093974\\
\texttt{1032} & -29505550692461 & 7125110669631 & \texttt{1125} & -721622838955625 & -1001800045025884\\
\texttt{1100} & 36584808817 & 4443292531 & \texttt{1205} & 954331738742774 & 86139946676354\\
\texttt{1101} & -384686522151 & -214267967816 & \texttt{1215} & -2967160593886730 & -1778279773878844\\
\texttt{1102} & 5324211335225 & 2725226108516 & \texttt{1225} & 230712195247103 & 1124739997585059\\
\texttt{1110} & -359366710405 & 17668350027 & \texttt{2025} & -301050856529182 & 85453661070936\\
\texttt{1111} & 2842945084676 & 1290605926543 & \texttt{2125} & 2027333686810472 & 1352300413695861\\
\texttt{1112} & -34958318261751 & -25932152712266 & \texttt{2205} & -1069447924993083 & -580239324017409\\
\texttt{1120} & 518559205479 & -570745475571 & \texttt{2215} & 609208442192807 & 938619080785187\\
\texttt{1121} & 4391081381635 & 6063277800597 & \texttt{2225} & -11011599217420063 & -6212585421173936\\
\texttt{1122} & -225020420660318 & -67920139935755 & \texttt{1006} & 41565372477041 & 26263841244901\\
\texttt{1130} & 2710012733037 & 1431234147579 & \texttt{1016} & -956197368394920 & -633775613129915\\
\texttt{1131} & 29514233383267 & -6322719497834 & \texttt{1026} & 803558607753700 & 1237489029357774\\
\texttt{1132} & 46176043952700 & 41276591428629 & \texttt{1106} & 408396291795931 & 265646865588706\\
\texttt{1200} & -121540878200 & -10543564096 & \texttt{1116} & -311545779339601 & 1000776388530746\\
\texttt{1201} & 119519012188 & -88285769173 & \texttt{1126} & 2217781576763753 & 285558633215349\\
\texttt{1202} & -6127237420960 & -2579576702179 & \texttt{1206} & -1079110326624621 & -1150476929513273\\
\texttt{1210} & 1992333428082 & 378293390911 & \texttt{1216} & 4281688133593783 & 3126632900574151\\
\texttt{1211} & -850875691279 & 2417350027267 & \texttt{1226} & -1768367748254345 & -2783877979138021\\
\texttt{1212} & 27404679667076 & -3048750830427 & \texttt{2006} & -1300104952969944 & -281682351519919\\
\texttt{1220} & -14121133101518 & -5352596754130 & \texttt{2016} & 3529116736263219 & 2200397679137744\\
\texttt{1221} & 99970272460171 & 45070182629416 & \texttt{2026} & -2335457336259735 & -1333938784093371\\
\texttt{1222} & 129485671547825 & 7930338677259 & \texttt{2106} & 835490613756248 & 1076867292897843\\
\texttt{1230} & -12001348728943 & 2470109848577 & \texttt{2116} & 3490784539223737 & 796920963025630\\
\texttt{1231} & -74662196352396 & -26850762639709 & \texttt{2126} & -1300245154128385 & -1205528217928385\\
\texttt{1232} & 7352501292907 & 4349039153086 & \texttt{2206} & 1545840549778553 & 587548223128652\\
\texttt{1300} & 139261157104 & 16003351818 & \texttt{2216} & -3617752634843183 & -2025347056910390\\
\texttt{1301} & -2251641203817 & -1815109126323 & \texttt{2226} & 1998959243044620 & 1141258341259932\\
\texttt{1302} & 29991868688401 & 20699864959570 & \texttt{3006} & 853857582634057 & 943222972985452\\
\texttt{1310} & 539465484461 & 1757190038586 & \texttt{3016} & 63307252255650 & -557105167452987\\
\texttt{1311} & -30209600878313 & -16000233686762 & \texttt{3106} & 159936527287197 & -254975093159509\\
\texttt{1312} & 35013599387107 & -24987398111253 & \texttt{3116} & -876255539770078 & -149925721629505\\
\texttt{1320} & 11192245339152 & -2984397699328 & \texttt{4005} & -174277712987940 & -120030259459457\\
\texttt{1321} & -118559349534260 & 4053950662716 & \texttt{4015} & 31481958479575 & 57470869235428\\
\texttt{1322} & 123011889380654 & 104054513240857 & \texttt{4105} & 70776106967460 & 98599257956683\\
\texttt{1330} & 35156955360678 & 2860190498108 & \texttt{4115} & -4278036026294 & -101012706931568\\
\texttt{1331} & -7598390144171 & -1831261178609 & \texttt{5000} & 20479606991 & 4882210231\\
\texttt{1332} & -13293446848633 & -1508450384613 & \texttt{5010} & -37788175571 & -10195383127\\
\texttt{2000} & 495910423 & 17332376 & \texttt{5100} & -39464675652 & -11659660379\\
\texttt{2001} & -32870288407 & 19711616023 & \texttt{5110} & 52979866480 & 32679575653\\
\texttt{2002} & 1037477959901 & 550492930809 & \texttt{5001} & -142793937105 & -44281298165\\
\texttt{2010} & -1666741075 & -74824379 & \texttt{5011} & 308637668804 & 82446249474\\
\texttt{2011} & -730246515484 & -626092019980 & \texttt{5101} & 295864712564 & 90582751561\\
\texttt{2012} & -27337596779318 & -12336081718050 & \texttt{5111} & -493619787609 & -163955545163\\
\texttt{2020} & 1850448937 & 93140950 & \texttt{4030} & -11569536069 & -4950645422\\
\texttt{2021} & 4618217620020 & 2020231167353 & \texttt{4130} & -548236696778 & -227789624986\\
\texttt{2022} & -12487582268497 & 5377544027838 & \texttt{4230} & -27839806998 & 708430594589\\
\texttt{2030} & -682009477 & -36515953 & \texttt{4300} & -122589585035 & -100902703390\\
\texttt{2031} & -2570600310043 & -1109832440731 & \texttt{4310} & -570521042151 & -272577926594\\
\texttt{2032} & 16402974539787 & 16752401466924 & \texttt{4320} & 52188912231 & 1080945561527\\
\texttt{2100} & -9039040122 & -23130663235 & \texttt{4330} & 578901110928 & -651464246814\\
\texttt{2101} & -1004604423464 & -711337785608 & \texttt{4031} & 431180037899 & 166145588213\\
\texttt{2102} & -23261438770928 & -11982016205480 & \texttt{4131} & 7340353039838 & 2059200500007\\
\texttt{2110} & 862032252413 & 638692185221 & \texttt{4231} & -7579098788653 & -4215722851719\\
\texttt{2111} & 29845576888129 & 15239314010842 & \texttt{4301} & 1927404524251 & 1435313109011\\
\texttt{2112} & -138502237647964 & -86073084064424 & \texttt{4311} & 4957098548815 & 244190407692\\
\texttt{2120} & -4753514034369 & -2033982281903 & \texttt{4321} & -6790797895836 & -5141039803014\\
\texttt{2121} & 8496717028518 & -6379782660626 & \texttt{4331} & -353635707153 & 2898517581638\\
\texttt{2122} & 148888378805789 & 208458038180378 & \texttt{4032} & -4517310483341 & -1399225933760\\
\texttt{2130} & 2616360191419 & 1114579885303 & \texttt{4132} & -21180169633809 & -6260890463694\\
\texttt{2131} & -17127109550523 & -30254173729145 & \texttt{4232} & 31935575616809 & 11461714137638\\
\texttt{2132} & -27824592631050 & -21412008103231 & \texttt{4302} & -11506225975055 & -7380195243786\\
\texttt{2200} & -60509433237 & 24093932383 & \texttt{4312} & -6535314442104 & 5336157353790\\
\texttt{2201} & 3974977148772 & 2079023581736 & \texttt{4322} & 24936933665051 & 8518725966391\\
\texttt{2202} & 29767815121315 & 8120116696411 & \texttt{4332} & -7283057531282 & -6193420474812\\
\texttt{2210} & -4079907542705 & -2956634148574 & \texttt{3026} & -250957585922224 & -207719233552\\
\texttt{2211} & -27986538369798 & 316574091051 & \texttt{3126} & 565905867262893 & 157651340686222\\
\texttt{2212} & 227329346119765 & 147839480222492 & \texttt{3206} & -379271226623685 & -187134297406105\\
\texttt{2220} & 6015507679793 & 1199897735188 & \texttt{1007} & -27914199162631 & -16004546324351\\
\texttt{2221} & -51483243818746 & -52626597463674 & \texttt{1017} & 922775940233152 & 632339296659259\\
\texttt{2222} & 77344465359048 & -85545942612860 & \texttt{1027} & -1112210345222203 & -1319518461596673\\
\texttt{2230} & 16481086143 & 13423675879107 & \texttt{1107} & -505684548727059 & -391387202354517\\
\texttt{2231} & 19804999972254 & 48300132264986 & \texttt{1117} & 212634396828437 & -1141231357515259\\
\texttt{2232} & 44213597070667 & -6231788936519 & \texttt{1127} & -2402175171448575 & 1590594425726518\\
\texttt{2300} & 14015104533 & -10506364396 & \texttt{1207} & 651063382097495 & 1246884102619553\\
\texttt{2301} & -9786288641389 & -6166234236214 & \texttt{1217} & -1790607320573011 & -1845480746575228\\
\texttt{2302} & 20701510003523 & 10914249819054 & \texttt{1227} & 2072017518704031 & 1166366337118542\\
\texttt{2310} & 9878878568464 & 6489015654447 & \texttt{1008} & 7565199678755 & 3986558221826\\
\texttt{2311} & 8360196172154 & 1181960735152 & \texttt{1018} & -494874079449964 & -254786572034162\\
\texttt{2312} & -118554422111671 & -69645372639132 & \texttt{1028} & 612778648206724 & 430165284755218\\
\texttt{2320} & -30869015307586 & -14908177857769 & \texttt{1108} & 372774484506249 & 191281684234430\\
\texttt{2321} & 103458106438401 & 100487898835959 & \texttt{1118} & 1027772764836381 & 597926805018560\\
\texttt{2322} & 1770135168584 & -25782310390277 & \texttt{1128} & -413281042385611 & -666673535642829\\
\texttt{2330} & 14157621256764 & -25201395069489 & \texttt{1208} & -1204271494610495 & -735162587366260\\
\texttt{2331} & -72585085237697 & -8205529524059 & \texttt{1218} & 775889431512947 & 628435991352363\\
\texttt{2332} & 47948897024248 & 30804540360387 & \texttt{1228} & -347986012578118 & -115989095853637\\
\texttt{3000} & 1472392103 & 2253170088 & \texttt{2007} & 913912453194274 & 19685095677415\\
\texttt{3001} & -537122585526 & -341215861554 & \texttt{2017} & -3990344614190265 & -2060275362648049\\
\texttt{3002} & -2325792906236 & -159749948089 & \texttt{2027} & 11780210884844241 & 7049975235082321\\
\texttt{3010} & 12218259209 & -9216011567 & \texttt{2107} & -623359245178004 & 336507905655200\\
\texttt{3011} & 1441882211612 & 1256833194490 & \texttt{2117} & 211977859844158 & -49099714168745\\
\texttt{3012} & -7435398779386 & -4512065531294 & \texttt{2127} & -75791922196404 & 444008879327699\\
\texttt{3020} & -12111685672 & 11021918552 & \texttt{2207} & -7691656739498755 & -4196439269155395\\
\texttt{3021} & -2316852442582 & -1717179714920 & \texttt{2217} & 1779660939528072 & 691111489167821\\
\texttt{3022} & 16571466238779 & 14855935348378 & \texttt{2227} & -890630997661623 & -609304434185255\\
\texttt{3030} & -1845359795 & -4550133873 & \texttt{2008} & -133584679831656 & -36671072853272\\
\texttt{3031} & 1199689972932 & 808832265211 & \texttt{2018} & 1636280679995619 & 955397430279777\\
\texttt{3032} & -1023642514930 & -1145015296450 & \texttt{2028} & -791122601713459 & -632228010254450\\
\texttt{3100} & 113682031531 & 63789575119 & \texttt{2108} & -412966344978939 & -432180461646892\\
\texttt{3101} & -366437665635 & 193046082304 & \texttt{2118} & -429051733953938 & -194095807498706\\
\texttt{3102} & 8810517364265 & 106718601591 & \texttt{2128} & 370381037208746 & 268507209657725\\
\texttt{3110} & 149570015490 & -216826377117 & \texttt{2208} & 261724034031444 & 319380440037321\\
\texttt{3111} & 1724364876685 & -1018308107636 & \texttt{2218} & -320328631596631 & -264471012888398\\
\texttt{3112} & 30502040372385 & 50986777488253 & \texttt{2228} & 93563287681798 & 80649684673528\\
\texttt{3120} & 822258797006 & 687842998287 & \texttt{3216} & 733486893902033 & 407221777036940\\
\texttt{3121} & -2061170778245 & -2400164800736 & \texttt{3226} & -315210425996834 & -167906954275334\\
\texttt{3122} & 43976540038336 & -4212054036803 & \texttt{3007} & 0 & 0\\
\texttt{3130} & -710546839847 & -493221025847 & \texttt{3017} & 548520204313533 & 446075088510153\\
\texttt{3131} & -10396074126278 & -8554107354880 & \texttt{3027} & 152371490733378 & 95135571792271\\
\texttt{3132} & 8953146684941 & 14208378192280 & \texttt{3107} & 948844320809248 & 696273652540970\\
\texttt{3200} & -133682110505 & -115902749191 & \texttt{3117} & -383777157066106 & -439723665176857\\
\texttt{3201} & -1332445760427 & -2044721131781 & \texttt{3127} & -233413609090982 & -70633107812082\\
\texttt{3202} & -14587022318338 & -1709406369229 & \texttt{3207} & -119020198861610 & -135693956518777\\
\texttt{3210} & -34167085078 & 1560761861102 & \texttt{3217} & -109855896007212 & 8932209144029\\
\texttt{3211} & 25722365357402 & 13248206833760 & \texttt{3227} & 145677094971316 & 59274296098146\\
\texttt{3212} & -27507165527671 & -33944873330226 & \texttt{3008} & 402027040291502 & 195854014548807\\
\texttt{3220} & -6785897643010 & -6832944302085 & \texttt{3018} & -278397278109365 & -190847542344792\\
\texttt{3221} & -30948360912729 & -1216159078332 & \texttt{3028} & 2301715248140 & -852223915750\\
\texttt{3222} & 54472467344717 & 42726057960336 & \texttt{3108} & -428153869054383 & -207374329008489\\
\texttt{3230} & 9270390138450 & 8007300598237 & \texttt{3118} & 334431460645505 & 224725879716749\\
\texttt{3231} & 3822682702244 & -5281117897295 & \texttt{3128} & -6786265843973 & -13626571207801\\
\texttt{3232} & 121698331042 & -2238735679314 & \texttt{3208} & 101125091507706 & 62653434960105\\
\texttt{3300} & 36305504889 & 52882605708 & \texttt{3218} & -75060092614671 & -65845034836410\\
\texttt{3301} & 2508322777133 & 1975704951535 & \texttt{3228} & -7342064584326 & 5831078329183\\
\texttt{3302} & -2230124471698 & -3182005477337 & \texttt{4033} & 13255138457562 & 4613876238618\\
\texttt{3310} & -1817393169414 & -1940311154249 & \texttt{4133} & 15205795902700 & 3626617091019\\
\texttt{3311} & -10128708748269 & -5287680721419 & \texttt{4233} & -37920958144578 & -10100391985055\\
\texttt{3312} & 3584252122957 & 15405566113760 & \texttt{4303} & 28295419901481 & 17610729115787\\
\texttt{3320} & 6753739426992 & 5914092275189 & \texttt{4313} & -17849541296200 & -23789850793002\\
\texttt{3321} & 20668924284157 & -1588120776891 & \texttt{4323} & -19995874880589 & 3059149630881\\
\texttt{3322} & -15230833680344 & -4603993136536 & \texttt{4333} & 12267627150248 & 4536450568087\\
\texttt{3330} & -11928752723824 & -7563656769662 & \texttt{4034} & -11191198047324 & -4242815780107\\
\texttt{3331} & 2781854080681 & 5941816991036 & \texttt{4134} & 4662164037556 & 3601912133557\\
\texttt{3332} & 1923541249589 & -1648440161819 & \texttt{4234} & 8053908751530 & -1652505884092\\
\texttt{4000} & -25058540291 & 3903956222 & \texttt{4304} & -23608192273932 & -14729698015088\\
\texttt{4001} & 676799999244 & 137428193721 & \texttt{4314} & 30577235196818 & 25898325111827\\
\texttt{4010} & 28972194979 & -12045903636 & \texttt{4324} & -6173093619767 & -13901948548640\\
\texttt{4011} & -882391506636 & -226153760731 & \texttt{4334} & -3239145990125 & 1215604221801\\
\texttt{4100} & -72097501508 & -94975232198 & \texttt{4025} & 38154900606683 & 4470357808257\\
\texttt{4101} & 816746868855 & 772950207428 & \texttt{4125} & -46198941331046 & 12362630627114\\
\texttt{4110} & -584816999275 & -20915338093 & \texttt{4205} & 22230916170662 & -18282716444558\\
\texttt{4111} & 5662790016589 & 1275205020827 & \texttt{4215} & -17515951332757 & 49147874487931\\
\texttt{1003} & -488161911927 & -232004586488 & \texttt{4225} & 9031948033218 & -21544281403358\\
\texttt{1013} & 24724963881974 & 11869874719390 & \texttt{4006} & 153267564599700 & 102899021742617\\
\texttt{1103} & -32713211753092 & -11884566764381 & \texttt{4016} & -97897610067683 & -73558288559717\\
\texttt{1113} & 325472743620466 & 197962826763849 & \texttt{4026} & 12959521367441 & 13870491874553\\
\texttt{2003} & 26439476322538 & 12090365830548 & \texttt{4106} & -124226700212963 & -103852300842524\\
\texttt{2013} & 163058050486085 & 80679842735976 & \texttt{4116} & 102527817248177 & 105056867757343\\
\texttt{2103} & 146239922912160 & 106902336623288 & \texttt{4126} & -14786113747372 & -25678121405337\\
\texttt{2113} & -131816471947232 & -306528751815631 & \texttt{4206} & 21082719577281 & 26239212155170\\
\texttt{1004} & 9256393919436 & 9087042117806 & \texttt{4216} & -25576070252432 & -41072615417184\\
\texttt{1014} & -216577890831676 & -135416267967323 & \texttt{4226} & 5573833076926 & 15227957711464\\
\texttt{1104} & 107675646684138 & 7505938741824 & \texttt{5020} & 17288792739 & 5172646801\\
\texttt{1114} & -584992838865985 & -119309168748518 & \texttt{5120} & -11751292083 & -20053907468\\
\texttt{2004} & -239539768695648 & -135610012700538 & \texttt{5200} & 19693690196 & 7759247552\\
\texttt{2014} & -93977517664715 & 157645515439953 & \texttt{5210} & -10861115810 & -31327432762\\
\texttt{2104} & -174173086808563 & -38736861592413 & \texttt{5220} & -15348222892 & 20294402314\\
\texttt{2114} & 840606793865061 & 1166326545787367 & \texttt{5021} & -159098268273 & -38164408423\\
\texttt{3003} & 36678946496680 & 16062770182332 & \texttt{5121} & 211769936465 & 76580307153\\
\texttt{3013} & -69278549264962 & -68582298378601 & \texttt{5201} & -154112317159 & -40614013160\\
\texttt{3103} & -133635425454777 & -100073352127505 & \texttt{5211} & 187276388577 & 104922719542\\
\texttt{3113} & -204499083472814 & -118688845321202 & \texttt{5221} & -6696058374 & -53814483199\\
\texttt{4002} & -7737826289373 & -1850240668425 & \texttt{5002} & 177094058022 & 44177096174\\
\texttt{4012} & 12531263790024 & 4117456701301 & \texttt{5012} & -667555121733 & -143865221252\\
\texttt{4102} & -243297367169 & -3524295409928 & \texttt{5022} & 416034717968 & 115055532444\\
\texttt{4112} & -5165617372041 & 4468919094092 & \texttt{5102} & -578393335561 & -182361374649\\
\texttt{1005} & -32503323207243 & -22387175164544 & \texttt{5112} & 1201195445502 & 337430254010\\
\texttt{1015} & 572882947081532 & 291365546632843 & \texttt{5122} & -667884374474 & -193546056581\\
\texttt{1105} & -224282868491517 & -6431611105840 & \texttt{5202} & 370552018332 & 113929742379\\
\texttt{1115} & 779268242449797 & -248907150474152 & \texttt{5212} & -599500468708 & -211439042548\\
\texttt{2005} & 715984817840254 & 377899568765353 & \texttt{5222} & 213962702851 & 103191920810\\
\texttt{2015} & -533263991585677 & -1048018799506979 & \texttt{5003} & -126475910662 & 8763689971\\
\texttt{2105} & -28496477099083 & -771381256913757 & \texttt{5013} & 455949985704 & 16690936911\\
\texttt{2115} & -3754759752831644 & -1513499714582479 & \texttt{5023} & -342483436016 & -101939949817\\
\texttt{3004} & -35538343375266 & 56575127499111 & \texttt{5103} & 384105862501 & 48231182047\\
\texttt{3014} & 382003434317316 & 195168507807285 & \texttt{5113} & -939870578097 & -201831097260\\
\texttt{3104} & 535687034796776 & 370073129539420 & \texttt{5123} & 598657040747 & 186565636211\\
\texttt{3114} & 183225254862284 & 26288158356799 & \texttt{5203} & -272020723478 & -79478527065\\
\texttt{4003} & 18387703771522 & -3539041879551 & \texttt{5213} & 507910790955 & 157370993560\\
\texttt{4013} & -61931352795809 & -15255400161081 & \texttt{5223} & -245019371704 & -91663794268\\
\texttt{4103} & -24010891688285 & 7469429039999 & \texttt{3035} & 25636715366886 & 16074554593114\\
\texttt{4113} & 15877227143396 & -29264253136880 & \texttt{3135} & -31588610147003 & -23541235673853\\
\texttt{4020} & 7884465198 & 13433256490 & \texttt{3235} & 6747692761451 & 9388557182747\\
\texttt{4200} & 218304084741 & 192085446514 & \texttt{3305} & 17106769614269 & -9956292323141\\
\texttt{4120} & 1160711311187 & 359181172143 & \texttt{3315} & -14139581940909 & 12826455725761\\
\texttt{4210} & 879404714583 & 148760169733 & \texttt{3325} & -743860053902 & -7427345349846\\
\texttt{4021} & -243998778723 & -107611398810 & \texttt{6000} & -146190766 & -92576825\\
\texttt{4201} & -2813918993768 & -2046748391829 & \texttt{6010} & 305025702 & 130801140\\
\texttt{4121} & -14150184026483 & -5029308114287 & \texttt{6020} & -195621876 & -37918993\\
\texttt{4211} & -8784453140929 & -1270855546287 & \texttt{6100} & 348830142 & 196297251\\
\texttt{4022} & 1051133682638 & 756333716488 & \texttt{6110} & -718617403 & -329102163\\
\texttt{4202} & 15630643364022 & 11496191831900 & \texttt{6120} & 455326383 & 146727957\\
\texttt{4122} & 32848970939367 & 9782619482920 & \texttt{6200} & -267207108 & -54780319\\
\texttt{4212} & 4274672164540 & -8582798662713 & \texttt{6210} & 491939029 & 127116753\\
\texttt{3023} & -27884371352994 & -7977275399647 & \texttt{6220} & -238059801 & -75445328\\
\texttt{3203} & 95032318802936 & 52954106826080 & \texttt{6001} & 176103112 & -110631273\\
\texttt{3123} & -41741353130175 & 6916528688619 & \texttt{6011} & -494543910 & 51715199\\
\texttt{3213} & -30524207973561 & -27088479388724 & \texttt{6021} & 294817665 & -13463731\\
\texttt{1023} & 94188766026169 & -28300126607113 & \texttt{6101} & -493604682 & 152982706\\
\texttt{1123} & -33370242377739 & -132152485006557 & \texttt{6111} & 1108046031 & 71966088\\
\texttt{1203} & 30065470493522 & 17906270100347 & \texttt{6121} & -749183999 & -141289116\\
\texttt{1213} & -391697634988441 & -328254919596941 & \texttt{6201} & 458466132 & -13644855\\
\texttt{1223} & -450856864222802 & 259547084164919 & \texttt{6211} & -846111609 & -120200872\\
\texttt{1024} & -25564222254736 & 96575081175680 & \texttt{6221} & 397128286 & 100326291\\
\end{longtable}
\endgroup

\Needspace{14\baselineskip}

\begin{longtable}[]{@{}lrr@{}}
\caption{Outward enclosures of the specified trial expressions. The
residual norm includes the signed bulk-contact term.}\tabularnewline
\toprule\noalign{}
Expression & Lower endpoint & Upper endpoint \\
\midrule\noalign{}
\endfirsthead
\toprule\noalign{}
Expression & Lower endpoint & Upper endpoint \\
\midrule\noalign{}
\endhead
\bottomrule\noalign{}
\endlastfoot
\(F(f_{\rm L})\) & 1.826694805763 & 1.826694805764 \\
\(F(v)\) & 1.826674684188 & 1.826674684189 \\
\(\eta(v)\) & 0.000959301158 & 0.000959301159 \\
\(F(v)+K\eta(v)/6\) & 1.826715689395 & 1.826715689396 \\
\end{longtable}

These evaluations give the outward enclosure
\(1.8266948\le\Theta\le1.8267157\). Independent deterministic
integration differentiates the complete combined trial in Cartesian
coordinates before forming its products and evaluates the same
expressions at two shape-quadrature orders. This checks the reductions
and normalizations; the enclosure itself follows from the exact finite
reductions with outward arithmetic. Under the two physical conditions
stated in Section S2.2 and Proposition 1 of the main article, the same
interval bounds \(c_3(0)\).

\begingroup\setlength{\parskip}{3pt plus 1pt}
\section{References}\label{references-supplement}

\noindent\begin{minipage}{\linewidth}
{[}1{]} N. J. Wagner and W. B. Russel, ``Nonequilibrium statistical mechanics of concentrated colloidal dispersions: Hard spheres in weak flows with many-body thermodynamic interactions,'' \emph{Physica A}, vol.~155, no. 3, pp.~475--518, 1989 (doi: 10.1016/0378-4371(89)90003-4).
\end{minipage}\par

\noindent\begin{minipage}{\linewidth}
{[}2{]} J. Bławzdziewicz and G. Szamel, ``Structure and rheology of semidilute suspension under shear,'' \emph{Phys. Rev.~E}, vol.~48, no. 6, pp.~4632--4636, 1993 (doi: 10.1103/PhysRevE.48.4632).
\end{minipage}\par

\noindent\begin{minipage}{\linewidth}
{[}3{]} J. G. Kirkwood, ``Statistical mechanics of fluid mixtures,'' \emph{J. Chem. Phys.}, vol.~3, no. 5, pp.~300--313, 1935 (doi: 10.1063/1.1749657).
\end{minipage}\par

\noindent\begin{minipage}{\linewidth}
{[}4{]} B. U. Felderhof and R. B. Jones, ``Linear response theory of the viscosity of suspensions of spherical Brownian particles,'' \emph{Physica A}, vol.~146, no. 3, pp.~417--432, 1987 (doi: 10.1016/0378-4371(87)90278-0).
\end{minipage}\par

\noindent\begin{minipage}{\linewidth}
{[}5{]} R. A. Lionberger, ``Rheology, structure, and diffusion in concentrated colloidal dispersions,'' Ph.D.~dissertation, Dept. Chem. Eng., Princeton Univ., Princeton, NJ, USA, 1996.
\end{minipage}\par

\noindent\begin{minipage}{\linewidth}
{[}6{]} R. A. Lionberger and W. B. Russel, ``Effectiveness of nonequilibrium closures for the many body forces in concentrated colloidal dispersions,'' \emph{J. Chem. Phys.}, vol.~106, no. 1, pp.~402--416, 1997 (doi: 10.1063/1.473029).
\end{minipage}\par

\noindent\begin{minipage}{\linewidth}
{[}7{]} R. A. Lionberger and W. B. Russel, ``Microscopic theories of the rheology of stable colloidal dispersions,'' in \emph{Advances in Chemical Physics}, vol.~111, I. Prigogine and S. A. Rice, Eds., New York, NY, USA, John Wiley \& Sons, 2000, pp.~399--474 (ISBN: 0-471-34990-9; doi: 10.1002/9780470141700.ch3).
\end{minipage}\par

\noindent\begin{minipage}{\linewidth}
{[}8{]} G. Szamel, ``Nonequilibrium structure and rheology of concentrated colloidal suspensions: Linear response,'' \emph{J. Chem. Phys.}, vol.~114, no. 19, pp.~8708--8717, 2001 (doi: 10.1063/1.1366335).
\end{minipage}\par

\noindent\begin{minipage}{\linewidth}
{[}9{]} E. Nazockdast and J. F. Morris, ``Microstructural theory and the rheology of concentrated colloidal suspensions,'' \emph{J. Fluid Mech.}, vol.~713, pp.~420--452, 2012 (doi: 10.1017/jfm.2012.467).
\end{minipage}\par

\noindent\begin{minipage}{\linewidth}
{[}10{]} T. Ohtsuki, ``Dynamical properties of strongly interacting Brownian particles. I. Dynamic shear viscosity,'' \emph{Physica A}, vol.~108, nos. 2--3, pp.~441--458, 1981 (doi: 10.1016/0378-4371(81)90141-2).
\end{minipage}\par

\noindent\begin{minipage}{\linewidth}
{[}11{]} S. M. Tschopp and J. M. Brader, ``Superadiabatic dynamical density functional theory for colloidal suspensions under homogeneous steady-shear,'' \emph{J. Chem. Phys.}, vol.~160, no. 21, 2024, Art. no. 214124 (doi: 10.1063/5.0211198).
\end{minipage}\par

\noindent\begin{minipage}{\linewidth}
{[}12{]} W. B. Russel and A. P. Gast, ``Nonequilibrium statistical mechanics of concentrated colloidal dispersions: Hard spheres in weak flows,'' \emph{J. Chem. Phys.}, vol.~84, no. 3, pp.~1815--1826, 1986 (doi: 10.1063/1.450428).
\end{minipage}\par

\noindent\begin{minipage}{\linewidth}
{[}13{]} J. F. Brady, ``The rheological behavior of concentrated colloidal dispersions,'' \emph{J. Chem. Phys.}, vol.~99, no. 1, pp.~567--581, 1993 (doi: 10.1063/1.465782).
\end{minipage}\par

\noindent\begin{minipage}{\linewidth}
{[}14{]} K. Scherwinski, ``Two approximations for the classical triplet distribution function,'' \emph{Mol. Phys.}, vol.~70, no. 5, pp.~797--809, 1990 (doi: 10.1080/00268979000101361).
\end{minipage}\par

\endgroup

\end{document}